\documentclass[12pt]{article}
\usepackage{epsf}
\usepackage{amsmath}
\usepackage{graphicx}
\usepackage{color}
\usepackage{psfrag}

\usepackage{ifpdf}

\newcommand{\bmat}{\left(\begin{array}}
\newcommand{\emat}{\end{array}\right)}

\def\yzero{\smash{\hbox{$y\kern-4pt\raise1pt\hbox{${}^\circ$}$}}}

\def\a{\alpha}
\def\b{\beta}

\def\beq{\begin{equation}}
\def\eeq{\end{equation}}
\def\beqa{\begin{eqnarray}}
\def\eeqa{\end{eqnarray}}

\def\-{\hphantom{-}}
\def\ov{\overline}
\def\s2{\frac{1}{\sqrt2}}

\def\beq{\begin{equation}}
\def\eeq{\end{equation}}
\def\beqa{\begin{eqnarray}}
\def\eeqa{\end{eqnarray}}
\def\tr{{\rm tr \,}}

\def\diag{{\rm diag \,}}
\def\IF{\relax{\rm I\kern-.18em F}}
\def\II{\relax{\rm I\kern-.18em I}}

\def\Dsl{\,\raise.15ex\hbox{/}\mkern-13.5mu D} 

\def\IZ{{\bf Z}}
\def\IX{{\bf X}}
\def\IT{{\bf T}}
\def\IP{\bf P}

\def\ti{\times}

\newcommand{\drawsquare}[2]{\hbox{%
\rule{#2pt}{#1pt}\hskip-#2pt
\rule{#1pt}{#2pt}\hskip-#1pt
\rule[#1pt]{#1pt}{#2pt}}\rule[#1pt]{#2pt}{#2pt}\hskip-#2pt
\rule{#2pt}{#1pt}}

\newcommand{\fund}{\raisebox{-.5pt}{\drawsquare{6.5}{0.4}}}
\newcommand{\Ysymm}{\raisebox{-.5pt}{\drawsquare{6.5}{0.4}}\hskip-0.4pt%
        \raisebox{-.5pt}{\drawsquare{6.5}{0.4}}}
\newcommand{\Yasymm}{\raisebox{-3.5pt}{\drawsquare{6.5}{0.4}}\hskip-6.9pt%
        \raisebox{3pt}{\drawsquare{6.5}{0.4}}}
\newcommand{\antifund}{\overline{\fund}}

\catcode`\@=11   
\newdimen\@rotdimen
\newbox\@rotbox  

\def\@vspec#1{\special{ps:#1}}
\def\@rotstart#1{\@vspec{gsave currentpoint currentpoint translate
   #1 neg exch neg exch translate}}
\def\@rotfinish{\@vspec{currentpoint grestore moveto}}
\def\@rotr#1{\@rotdimen=\ht#1\advance\@rotdimen by\dp#1%
   \hbox to\@rotdimen{\hskip\ht#1\vbox to\wd#1{\@rotstart{90 rotate}%
   \box#1\vss}\hss}\@rotfinish}
\def\@rotl#1{\@rotdimen=\ht#1\advance\@rotdimen by\dp#1%
   \hbox to\@rotdimen{\vbox to\wd#1{\vskip\wd#1\@rotstart{270 rotate}%
   \box#1\vss}\hss}\@rotfinish}%
\def\@rotu#1{\@rotdimen=\ht#1\advance\@rotdimen by\dp#1%
   \hbox to\wd#1{\hskip\wd#1\vbox to\@rotdimen{\vskip\@rotdimen
   \@rotstart{-1 dup scale}\box#1\vss}\hss}\@rotfinish}%
\def\@rotf#1{\hbox to\wd#1{\hskip\wd#1\@rotstart{-1 1 scale}%
   \box#1\hss}\@rotfinish}%
\def\rotate{\@ifnextchar[{\@rotate}{\@rotate[l]}}
\def\@rotate[#1]#2{\setbox\@rotbox=\hbox{#2}\@nameuse{@rot#1}\@rotbox}

\catcode`\@=12

\topmargin
-1.5cm
\textwidth
15.5cm
\textheight
23.5cm
\oddsidemargin
0.7cm
\evensidemargin
1.2cm

\begin{document}

\makeatletter
\@addtoreset{equation}{section}
\makeatother
\renewcommand{\theequation}{\thesection.\arabic{equation}}
\pagestyle{empty}
\rightline{ IFT-UAM/CSIC-11-42}
\vspace{0.1cm}
\begin{center}
\LARGE{\bf Discrete gauge symmetries\\ in D-brane models
\\[12mm]}
\normalsize{Mikel Berasaluce-Gonz\'alez$^{1,2}$, Luis E. Ib\'a\~nez$^{1,2}$, Pablo Soler$^{1,2}$, Angel M. Uranga$^2$\\[3mm]}
\footnotesize{${}^{1}$ Departamento de F\'{\i}sica Te\'orica,\\[-0.3em] 
Universidad Aut\'onoma de Madrid, 28049 Madrid\\
${}^2$ Instituto de F\'{\i}sica Te\'orica IFT-UAM/CSIC,\\[-0.3em] 
C/ Nicol\'as Cabrera 13-15, Universidad Aut\'onoma de Madrid, 28049 Madrid, Spain} \\[2mm] 
{\tt mikel.berasaluce@estudiante.uam.es, luis.ibannez@uam.es}\\ {\tt pablo.soler@uam.es, angel.uranga@uam.es}

\vspace*{2cm}

{\large{\bf Abstract}} \\[10mm]
\end{center}

{\small 
In particle physics model building discrete symmetries are often invoked to
forbid unwanted or dangerous couplings. A classical example is the R-parity of the 
MSSM, which guarantees the absence of dimension four baryon- and lepton-number
violating operators.  Although phenomenologically useful, these discrete symmetries are, in the context of field theory, poorly motivated at a more fundamental level. Moreover,  discrete {\em global} symmetries are expected to be violated in consistent couplings to quantum gravity, while their {\em gauged} versions are expected to actually exist. In this paper we study discrete gauge symmetries in brane models in string theory, and argue that they are fairly generic in this framework. In particular  we study the appearance of discrete gauge symmetries in  (MS)SM  brane constructions in string theory, and show that a few  discrete ${\IZ_N}$ gauge  symmetries, including  R-parity and {\it baryon triality},   appear naturally as remnants of continuous $U(1)$ gauge symmetries with St\"uckelberg $N(B\wedge F)$ couplings. Interestingly,  they correspond to
 the simplest anomaly-free discrete  symmetries of the MSSM as classified in the early 90's.
 We provide a number of examples based on  type IIA intersecting brane constructions 
 with a (MS)SM spectrum. We also study the appearance of discrete generalizations of
 R-parity in unified $SU(5)$ type IIA orientifolds and local F-theory $SU(5)$ GUTs.
 }

%
\newpage
\setcounter{page}{1}
\pagestyle{plain}
\renewcommand{\thefootnote}{\arabic{footnote}}
\setcounter{footnote}{0}

\vspace*{1cm}

\section{Introduction}
\label{intro}

Discrete symmetries are often invoked in particle physics model building in order to
forbid unwanted  terms in the Lagrangian. They have been used for example in order to
guarantee the absence of flavour changing neutral currents (FCNC) in two-Higgs models or
 in flavour models of fermion masses. In the context of the Minimal Supersymmetric Standard Model (MSSM), discrete symmetries seem unavoidable in order to explain the observed baryon stability. 
Indeed, a crucial difference between  the non-SUSY SM and the MSSM is that in the
latter the most general dimension four effective Lagrangian respects
neither baryon- nor lepton-number conservation.  The most general  superpotential consistent with gauge invariance and leading to dimension four operators has the structure
\beqa
W_{\rm MSSM}\, &=& \, Y_U^{ij}\, Q_iU_jH_u \,+\,Y_D^{ij}\,Q_iD_jH_d \,+\,Y_L^{ij}\,L_iE_jH_d \,+\,\mu \, H_uH_d \ +\nonumber\\ 
   & + & \,\lambda^{ijk}\, U_iD_jD_k \,+\, {\lambda^{ijk}}' \,Q_iD_jL_k\,+\,  {\lambda^{ijk}}'' \,L_iL_jE_k \,+\,\mu_R^i \,L_iH_u  \ 
\label{MSSMsuper}
\eeqa
where we use a standard notation for quark, lepton and Higgs superfields.
The first line contains the usual Yukawa couplings and the $\mu$-term, and respects baryon and lepton number; the $UDD$ terms in the second line violate baryon-number in one unit, whereas the rest violate lepton-number also in one unit. If all the terms in the second line were  present and unsuppressed,  the proton would decay with a lifetime of a few minutes.  The simplest solution to avoid this problem is to assume some discrete symmetry, like e.g. 
R-parity or baryon triality $B_3$, forbidding  all or some of the couplings in the second line.

Although indeed such discrete symmetries do their phenomenological job, their  fundamental origin is obscure. There are diverse arguments strongly suggesting that global symmetries, either continuous or discrete, are expected to be broken by quantum gravitational effects, and hence cannot exist in any consistent quantum theory including gravity (see \cite{Banks:1988yz,Abbott:1989jw,Coleman:1989zu} for early viewpoints, and e.g.\cite{Kallosh:1995hi,Banks:2010zn} and references therein, for more recent discussions). This suggests that discrete symmetries should have a gauge nature so that they are respected by such corrections \cite{Alford:1988sj,Krauss:1988zc,Preskill:1990bm}.
In particular ${\IZ_N}$ gauge symmetries may appear as discrete remnants of continuous $U(1)$ gauge 
symmetries when the latter are spontaneously broken by scalars with charge $N$, with other fields in the spectrum having charges not multiple of $N$.  Similarly to continuous gauge symmetries, 
 discrete gauge symmetries should respect certain anomaly cancellation conditions, which strongly restrict the possibilities in specific theories \cite{Ibanez:1991hv}.

In the case of the MSSM one can classify \cite{Ibanez:1991pr} the discrete gauge symmetries in terms of three discrete generators $R_N$, $A_N$, $L_N$ (see table \ref{generators}). Discrete anomaly cancellation  further constrains the possibilities, and the simplest
anomaly free discrete gauge symmetries are $R_2$ (which may be identified with the standard R-parity),
$B_3=R_3L_3$ (which is known as baryon triality), $L_3$ and $R_3L_3^2$ \cite{Ibanez:1991pr} 
(see also \cite{Banks:1991xj,Ibanez:1992ji,Martin:1992mq,Kurosawa:2001iq,Dreiner:2005rd,Mohapatra:2007vd,Araki:2008ek,Lee:2010gv,Kappl:2010yu} and references therein).
Each different symmetry forbids different
combinations of operators in the second line of  (\ref{MSSMsuper}). 
In the case of R-parity, the discrete gauge symmetry may be obtained as a ${\IZ_2}$ subgroup of a $U(1)_{B-L}$ gauge factor \cite{Font:1989ai}, typical of left-right symmetric extensions of the (MS)SM including right-handed 
neutrinos.  In the other cases the corresponding continuous $U(1)$ symmetries are less attractive and
include the introduction of new exotic charged particles; the discrete versions are however perfectly consistent with the minimal MSSM content.

If string theory is to describe the observed physics, a natural question  is whether discrete gauge symmetries arise in string compactifications.
Unlike in field theory, in string theory  constructions  we are not free to impose any symmetry, rather one should determine whether they are present or not in each given model. In the context of e.g.  heterotic orbifold or free fermion semi-realistic constructions there are typically a number of $U(1)$ gauge symmetries beyond hypercharge,
one of which may be identified with $U(1)_{B-L}$.  In principle one may obtain R-parity or other discrete gauge symmetry
by taking D- and F-flat directions in the scalar potential in which an appropriate scalar with charge $N$ (charge 2 in the case of R-parity) gets a vev (see e.g. ref.\cite{Lebedev:2007hv} for an attempt in this direction).  However  this mechanism is very much dependent on the existence and dynamical 
preference for  a particular choice of flat direction. The assessment of the existence of the symmetry thus requires a delicate analysis of this point. We would rather like to know whether there is a mechanism within string theory by which interesting discrete gauge symmetries survive in a natural way, without tuning scalar vevs  to that purpose.

In this paper we show that physically interesting discrete gauge symmetries are generic in large
classes of string compactifications. In particular type II orientifold constructions contain $U(1)_{\alpha}$ symmetries on the worldvolume of D-branes, which are generically broken to ${\IZ_N}$ discrete gauge symmetries by the presence of $N\, B\wedge F_{\alpha}$ couplings, with $B$ being Ramond-Ramond (RR) 2-form fields. This is a stringy implementation of the mechanism in  \cite{Banks:2010zn}\footnote{For a formal description in terms of stacks, see \cite{Hellerman:2010fv}.} (see also \cite{Camara:2011jg}). Such couplings are pervasive in explicit D-brane models: for anomalous $U(1)$'s they are required for the Green-Schwarz anomaly cancellation mechanism, and they also play an important role in removing additional $U(1)$'s (either anomalous or not) beyond hypercharge from the massless spectrum. The $U(1)$ symmetries remain exact in perturbation theory, but are violated by D-brane instanton effects, which may be (and are in fact often claimed to be) important in certain mod
 els. The point of this paper is the analysis of the exact discrete gauge remnants of
  these symmetries. We show how the resulting 
discrete gauge symmetries are free of mixed gauge and gravitational anomalies. We also present type IIA intersecting D6-brane examples with  spectrum close to the (MS)SM, and show that the required couplings  indeed appear and discrete gauge symmetries survive. Interestingly, these discrete gauge symmetries precisely correspond to the  anomaly free  $\IZ_2$, $\IZ_3$ symmetries described in the early 90's    \cite{Ibanez:1991pr} (see also \cite{Dreiner:2005rd}) or combinations thereof. In particular, R-parity and baryon triality often appear as discrete gauge symmetries of the effective actions. We also discuss how in unified $SU(5)$ orientifold or F-theory models R-parity and the corresponding $R_N$ generalizations may appear. 

The outline of this paper is as follows.
In section \ref{general-analysis} we describe the origin of discrete gauge symmetries from $B\wedge F$ couplings  in type IIA orientifolds with intersecting D6-branes. We  show that D-brane instantons preserve the discrete gauge symmetries, which are also shown to be automatically anomaly free.
In section \ref{sec:discrete-sm} we focus on D-brane (MS)SM brane constructions. In section \ref{ir-symmetries} we  review the classification of anomaly free discrete gauge
symmetries in the MSSM as formulated in \cite{Ibanez:1991pr}; in section \ref{sm-brane-symmetries} we describe their embedding in general D-brane models realizing the (MS)SM spectrum, and display explicit toroidal orientifold models in which these discrete gauge symmetries appear naturally.  Only a few such symmetries, including R-parity and baryon triality, actually appear, in agreement with \cite{Ibanez:1991pr}.
We also discuss, in section \ref{sec:su5}, that in the case of D-brane models with a  $SU(5)$ symmetry, the possibilities of  discrete gauge symmetries are much more restricted.
The appearance of discrete gauge symmetries in  local $SU(5)$ F-theory models, and in particular the realization of generalized R-parities, is addressed in section \ref{sec:fth}. In section \ref{k-theory} we comment on $\IZ_2$ discrete gauge symmetries associated to the discrete K-theory charge cancellation conditions, and suggest the intriguing possibility of identifying it with R-parity  in explicit constructions. We present some further comments and  conclusions in chapter 6. 

We include  two appendices.
In the first  we briefly discuss some technical aspects of toroidal orientifolds with tilted tori. Finally, in appendix B we comment on the appearance of certain $\IZ_2$ discrete gauge symmetries (including R-parity in concrete examples) associated to sectors of D-brane instantons with $Sp$-type orientifold projection.

\section{Discrete gauge symmetries and D-branes}
\label{general-analysis}

In this section we describe the appearance of discrete gauge symmetries in D-brane models from the analysis of their $BF$ couplings. For concreteness we center on type IIA compactifications with D6-branes wrapped on intersecting 3-cycles (see \cite{d6mb,thebook} for reviews, and \cite{Blumenhagen:2000fp,Aldazabal:2000dg,Aldazabal:2000cn} for some of the original references); the results apply similarly to other constructions, like type IIB models with magnetized D-branes, D-branes at singularities, etc, (see \cite{thebook} for reviews) as expected from mirror symmetry. Also, they should admit a lift to M-theory on $G_2$ manifolds, along the lines of \cite{Camara:2011jg}.

\subsection{Discrete gauge symmetries from $BF$ couplings}
\label{bf-general}

We focus on orientifolds of type IIA on a CY $\IX_6$, with an orientifold action $\Omega {\cal R}(-1)^{F_L}$. Here ${\cal R}$ is an antiholomorphic involution of $\IX_6$, acting as $z_i\to {\ov z}_i$ on local complex coordinates, so it introduces O6-planes. The compactification also contains stacks of $N_A$ D6$_A$-branes wrapped on 3-cycles $\Pi_A$ (along with their orientifold images on $\Pi_{A'}$). We need not impose the supersymmetry conditions at this level, since the analysis  is essentially topological, and holds even in non-supersymmetric models. 

We introduce a basis of 3-cycles $\{\alpha_k\}, \{\beta_k\}$, even and odd under the geometric action ${\cal R}$,  with $k=1,\ldots, h_{2,1}+1$. We assume for simplicity that $\alpha_k\cdot \beta_{l}=\delta_{kl}$. An alternative class of orientifold actions, satisfying instead $\alpha_k\cdot \beta_l=2\delta_{kl}$ for some values of $k$, leads to very similar physical results, but requires a careful tracking of factors of 2; we relegate it to appendix \ref{ap:tilted-orientifolds}. We expand the wrapped cycles in this basis as
\beqa
\Pi_A\,=\, \sum_k \, (\, r_A^k \alpha_k\, +\, s_A^k\beta_k\, )\quad ,\quad
\Pi_{A'}\,=\, \sum_k \, (\, r_A^k \alpha_k\, -\, s_A^k\beta_k\, )
\eeqa
The RR tadpole cancellation conditions are
\beqa
\sum_A\, N_A\,[\Pi_A]\,+\,\sum_{A'}\, N_A\, [\Pi_{A'}]\,-4\, [\Pi_{\rm O6}]\, =\, 0
\label{rr-tadpoles}
\eeqa
where $[\Pi_{\rm O6}]$ denotes the total homology class of the O6-planes (with the $-4$ from their  RR charge, assumed to be negative). In addition to the above constraint, there are certain discrete K-theory charge cancellation conditions \cite{Uranga:2000xp}, which actually play an interesting role, discussed in Section \ref{k-theory}.

The chiral part of the spectrum is 
\beqa
{\rm Gauge ~ group} \quad & \prod_A \, U(N_A)\nonumber \\
{\rm Ch. ~ fermions}\quad & \sum_{AB} \,I_{AB}\, (\fund_A,\antifund_B)\, +\, \sum_{AB'} \,I_{AB'}\, (\fund_A,\fund_B)\, +\nonumber \\
& +\, \sum_A\,\left(\,  \frac{I_{AA'}+I_{A,{\rm O6}}}{2}\, \Yasymm_{A}\, +\,  \frac{I_{AA'}-I_{A,{\rm O6}}}{2}\, \Ysymm_{A}\, \right)
\eeqa
where $I_{AB}=[\Pi_A]\cdot [\Pi_B]$, $I_{AB'}=[\Pi_A]\cdot [\Pi_{B'}]$ and $I_{A,{\rm O6}}=[\Pi_A]\cdot [\Pi_{\rm O6}]$ are the relevant intersection 
numbers giving the multiplicities.

Since the RR 5- and 3-form are intrinsically odd and even under the orientifold, the KK reduction leads to the following basis of RR 2-forms and their dual RR scalars
\beqa
B_k\, =\, \int_{\beta_k} C_5\quad ,\quad a_k\,=\,\int_{\alpha_k}C_3\; , \quad \mbox{with $dB_k=*_{4d}da_k$}
\eeqa
The KK reduction of the D6-brane Chern-Simons action leads to the following $BF$ couplings
\beqa
S_{BF_A}\, =\, \frac 12\, \left(\, \int_{\Pi_A} C_5\wedge \tr F_A\, -\,\int_{\Pi_{A'}} C_5\wedge \tr F_{A}\,\right)\, =\, \sum_k N_A\,s_A^k\, B_k\wedge F_A 
\eeqa
where the factor of $1/2$ is due to the orientifold action, and the relative minus sign of the orientifold image contributions arises because $F_{A'}=-F_A$. Also, the factor of $N_A$ arises from the $U(1)_A$ trace normalization.

In general, the factor of $N_A$ implies the appearance of a $\IZ_{N_A}$ discrete gauge symmetry. This corresponds to the general fact that the actual gauge group on a stack of $N$ D-branes is $[SU(N)\times U(1)]/\IZ_N$, with the $\IZ_N$ corresponding to the center of $SU(N)$, i.e. the $N$-ality. Namely, the group element $\diag (\alpha,\ldots,\alpha)$ with $\alpha=e^{2\pi i/N}$ can be regarded as belonging to $SU(N)$ or to the diagonal $U(1)$; the quotient by $\IZ_N$ implies that the two possibilities should be regarded as completely equivalent. The charges of fields under this $\IZ_N$ are given by their $N$-ality, and so this $\IZ_N$ does not imply any selection rule beyond $SU(N)$ gauge invariance; hence, it is not very interesting by itself.

The structure of the above coupling shows that an additional $\IZ_n$ symmetry appears when the coefficients $s_A^k$ are multiples of $n$, for all $k$; more precisely, when $n={\rm gcd}(s^k_A)$. In general, we may be interested in discrete subgroups of $U(1)$ linear combinations of the form
\beqa
Q\, =\, \sum_A \,c_A\, Q_A
\label{lincomb}
\eeqa
In order to properly identify the discrete gauge symmetry from the $BF$ coupling, we fix the normalization such that $c_A\in \IZ$, and ${\rm gcd}(c_A)=1$. The $BF$ couplings read
\beqa
S_{BF}\, =\, \bigl(\,{\textstyle \sum_A} \, c_A N_A\, s_A^k\, \bigr)\, B_k\wedge F
\label{bf-gen}
\eeqa
where $F$ is the field strength associated to the $Q$ generator. 
So there is a $\IZ_n$ gauge symmetry if the quantities $(\sum_A c_A N_As_A^k)$ are multiples of $n$, for all $k$. 
 
In our normalization, fields in the fundamental of $SU(N_A)$ have $U(1)_A$ charges $q_A=1$, while fields in the two-index symmetric or antisymmetric tensor representation have $q_A=2$ (and the opposite charges for the conjugate representations). For a field with charges $q_A$ under the $U(1)_A$, its charge under the $\IZ_n$ symmetry is $\sum_A c_Aq_A$ mod $n$. 

For future convenience, we describe the action of the symmetry on the RR scalars dual to the 2-forms $B_k$. Under a $U(1)$ gauge transformation, the scalars $a_k$ shift as
\beqa
A_\mu\, \to\, A_\mu\,+\,\partial_\mu \lambda\quad \rightarrow\quad
a_k\,\to\, a_k \, +\, \sum_A c_A N_A s_A^k \lambda \, .
\label{gauge-transf}
\eeqa
We conclude by rewriting the condition for a $\IZ_n$ symmetry as
\beqa
\sum_A c_A N_A [\Pi_A]\cdot [\alpha_k]\, =\, 0 \; \mbox{ mod $n$, for all $k$} \, .
\label{zn-condition}
\eeqa
Although we have derived it in the situation where $[\alpha_k]\cdot[\beta_l]=\delta_{kl}$, this expression for the condition is valid even in cases where $[\alpha_k]\cdot[\beta_l]=2\delta_{kl}$ for some subset of the $k$'s, see appendix \ref{ap:tilted-orientifolds}.

It is important to emphasize that the $U(1)_A$ symmetries behave as exact global symmetries at the perturbative level. However, they are violated by non-perturbative effects, in particular D-brane instantons \cite{Blumenhagen:2006xt,Ibanez:2006da,Florea:2006si} (see \cite{Blumenhagen:2009qh,Cvetic:2011vz,thebook} for reviews). 
The existence of a gauged discrete subgroup implies that it will be preserved by any such non-perturbative effect, as we describe more explicitly in section \ref{sec:instanton-effects}. One may think that for practical purposes, instanton effects may be negligible, and discrete gauge symmetries are irrelevant, since they are just part of the perturbatively exact global symmetries. However, in many SM-like D-brane models, instanton effects are often invoked to generate phenomenologically interesting (but perturbatively forbidden) couplings, see e.g. \cite{Blumenhagen:2007zk,Cvetic:2007qj,Cvetic:2007ku,Ibanez:2008my}, and so must be non-negligible. Hence it is relevant to ensure that other instantons do not induce dangerous couplings. Discrete gauge symmetries are an efficient way to guarantee this property.

\subsection{D-brane instanton effects}
\label{sec:instanton-effects}

Type IIA compactifications have non-perturbative effects from D2-brane instantons on 3-cycles. Let us denote $\Pi_{\rm inst}$ the 3-cycle wrapped by the instanton (and possibly its orientifold image, if it wraps a 3-cycle not invariant under ${\cal R}$). Such $\Pi_{\rm inst}$ can be expanded in terms of the 3-cycles $\{\alpha_k\}$ as 
\beqa
[\Pi_{\rm inst}]\, =\, \sum_k r_{\rm inst}^k\,\alpha_k
\eeqa
In supersymmetric models, there are certain conditions for such instantons to contribute to the superpotential; instantons not satisfying them contribute to other higher-dimensional operators, and are often neglected. However, here we are interested in showing that {\em all} instantons respect the discrete gauge symmetries, hence we must not restrict to superpotential generating instantons, and not even to BPS instantons. Hence we must consider instantons in the most general possible class.

The non-perturbative contribution of the instanton to the 4d effective action contains a piece
\beqa
e^{-S_{\rm cl.}}\, =\, e^{-\frac{V}{g_s}\, +i \, a}
\eeqa
where
\beqa
a= \int_{\Pi_{\rm inst}} C_3\, =\, \sum_k \, r_{\rm inst}^k\, a_k
\eeqa
Hence, under a $U(1)$ gauge transformation (\ref{gauge-transf}), the instanton exponential rotates by a phase
\beqa
 \sum_k \, r_{\rm inst}^k\,  \sum_A c_A N_A s_A^k \lambda
 \label{phase-shift}
\eeqa
As described in \cite{Blumenhagen:2006xt,Ibanez:2006da,Florea:2006si}, this phase rotation is cancelled by the insertion, in the complete instanton amplitude, of 4d fields charged under the $U(1)$ symmetry. This effectively leads to operators whose appearance was forbidden in perturbation theory.
Now in the presence of a discrete $\IZ_n$ gauge symmetry, namely when the quantities $(\sum_A c_A N_A s_A^k)$ are multiples of $n$ for all $k$, the instanton exponential shift is a multiple of $n$, so the non-perturbative effects preserve the discrete $\IZ_n$ subgroup. Conversely, the set of charged operators required to cancel the phase rotation of $e^{-S_{\rm cl.}}$ have $U(1)$ charges adding up to a multiple of $n$.

It is interesting to provide an alternative microscopic view of the argument. The  phase shift (\ref{phase-shift}) of the instanton exponent may be written as
\beqa
\sum_A c_A N_A \, \sum_k \, r_{\rm inst}^k\,  s_A^k \, =\,- \sum_A c_A N_A \, [\Pi_A] \cdot [\Pi_{\rm inst}]\, \equiv\,  -[\Pi_{Q}]\cdot [\Pi_{\rm inst}]
\eeqa
where in the first equality we have used $[\Pi_A]\cdot [\Pi_{\rm inst}]=\sum_k s_A^k \,r_{\rm inst}^l [\beta_k]\cdot [\alpha_l]=- \sum_k \, r_{\rm inst}^k\,  s_A^k$, and in the second we have defined $[\Pi_Q]=\sum_A c_AN_A [\Pi_A]$. The $\IZ_n$ discrete gauge symmetry implies that the intersection number of any instanton with the homology class associated to the $U(1)$ is multiple of $n$, as follows directly from \eqref{zn-condition}. This intersection number determines the number of instanton fermion zero modes charged under $U(1)$, and therefore the amount of $U(1)$ charge violation.

Let us finally remark on a complementary mechanism, already mentioned in \cite{Ibanez:2007rs}, to ensure that instantons preserve  discrete (presumably gauge) $\IZ_2$ symmetries. In models where all instantons mapped to themselves under the orientifold action experience an $Sp$ type orientifold projection (i.e. $\gamma_{\Omega}^2=-1$ for open strings with both endpoints on the instanton D-brane), the instanton class $\Pi_{\rm inst}$ expands in the basis $\alpha_k$ as a linear combination with {\em even} coefficients; in other words, the minimal instanton has worldvolume gauge group $USp(2)$, and arises from two D-brane instantons in the covering space. Hence, the violation of {\em any} $U(1)$ symmetry by instantons automatically preserves a $\IZ_2$ subgroup. A milder version guaranteeing a $\IZ_2$ subgroup of {\em some} $U(1)$, is that any instanton  intersecting the class $[\Pi_Q]$ of the $U(1)$ and invariant under the orientifold, is of $USp(2)$ type.  In appendix 
 \ref{ap:sp} we develop the realization of such $\IZ_2$ symmetries in a few examples, including a realization of R-parity in an SM-like D-brane construction. We also explain that these are discrete gauge symmetries, which can be made manifest in terms of the corresponding $BF$ couplings.

\subsection{Discrete anomaly cancellation}
\label{sec:anomaly}

The fact that all D-brane instantons (including gauge instantons) preserve these $\IZ_n$ symmetries suggests that they are anomaly-free (even if the corresponding $U(1)$'s are anomalous). It is worthwhile to verify this directly, using the conditions in \cite{Ibanez:1991hv}. 

Recall that states with charge $q_A$ under $U(1)_A$ have charge $\sum_A c_A q_A$ under the linear combination $Q$, and hence the same charge (mod $n$) under its $\IZ_n$ subgroup.

The mixed $\IZ_n-SU(N_B)^2$ anomaly is 
\beqa
&&\sum_A \, c_A\,N_A\, \frac 12 (I_{AB}+I_{AB'})=\frac 12 \sum_A \,c_A\,N_A\, [\Pi_A]\cdot([\Pi_B]+[\Pi_{B'}])=\nonumber \\
&&= \frac 12\sum_k  2 r_B^k   \sum_A \,c_A \,N_A\, [\Pi_A]\cdot [\alpha_k]
\eeqa
where the factor of $\frac 12$ arises from the $SU(N_B)$ quadratic Casimir in the fundamental. Using (\ref{zn-condition}), the above expression is of the form $\frac 12 n$ times an integer, as required by anomaly cancellation.

For the mixed $\IZ_n$-gravitational anomaly, we have
\beqa
&&\sum_A c_A\, \left\{\,  \sum_{B}\, N_AN_B \,I_{AB}\,+ \sum_{B'\neq A'} \,N_AN_B\,  I_{AB'}\, +\, 2\, \frac{I_{AA'}+I_{A,{\rm O6}}}{2} \, \frac{N_A(N_A-1)}{2}\, +\right. \nonumber \\
&&\left. \quad +\, 2\, \frac{I_{AA'}-I_{A,{\rm O6}}}{2}\, \frac{N_A(N_A+1)}{2} \, \right\} \, =\\
&& =\, \sum_A c_A\, \left(\, \sum_{B}\, N_AN_B \,I_{AB}+ \sum_{B'} N_AN_B I_{AB'}-N_AI_{A,{\rm O6}}\,\right)\,  =\, 3\, \sum_A c_AN_A\,I_{A,{\rm O6}}\nonumber
\eeqa
where in the last line, the sum in $B'$ includes $A'$, and in the last equality we use the tadpole condition (\ref{rr-tadpoles}). Since $[\Pi_{\rm O6}]$ must be an integer linear combination of the 3-cycles $[\alpha_k]$, the condition (\ref{zn-condition}) ensures that the last expression is a multiple of $n$, as required by anomaly cancellation.

The cancellation of mixed anomalies with other $U(1)$'s proceeds in an analogous fashion. The cubic ${\IZ_N^3}$ anomalies on the other hand cancel as an automatic 
consequence of the $U(1)^3$ anomaly cancellation in this setting.

It is interesting to compare the situation with discrete gauge symmetries in heterotic compactification, studied mostly in the context of toroidal orbifolds. Discrete gauge symmetries arise from continuous $U(1)$'s broken by vevs of dynamical fields with charge $n$. If the $U(1)$ is the (unique) anomalous one, it is possible to generate anomalous discrete gauge symmetries \cite{Ibanez:1992ji}, with anomaly canceled by the Green-Schwarz mechanism, as for the parent $U(1)$. The physical interest of this situation leans on the fact that instantons violating the discrete symmetry are necessarily very much suppressed, since their strength is controlled by SM gauge couplings. In our present D-brane constructions, we are implicitly focusing on symmetries preserved by {\em any} instanton, which are hence non-anomalous. Still, there are situations in which it may be physically meaningful to relax this requirement and consider anomalous discrete symmetries. For instance, in models wher
 e a subset of instantons have large strength (e.g. to generate neutrino masses or Yukawa couplings), while the remaining are hierarchically suppressed in comparison. Then, discrete symmetries respected by the former and violated by the latter could be anomalous, and behave similarly to the above mentioned heterotic ones.

\subsection{Toroidal orientifolds}
\label{toroidal-general}

In this section we particularize the above general analysis to the case of toroidal orientifolds. This is also valid for orbifolds thereof, as long as the relevant D6-branes do not wrap twisted cycles; this will be the case in the $\IZ_2\times \IZ_2$ orbifolds in the examples in the next section.

Consider a $\IT^6$, taken factorizable for simplicity, with each $(\IT^2)^i$ parametrized by $x^i$, $y^i$, $i=1,2,3$, and denote $[a_i]$, $[b_i]$ the 1-cycles along its two independent 1-cycles (with $[a_i]\cdot[b_j]=\delta_{ij}$). The orientifold acts as $x^i\to x^i$, $y^i\to -y^i$, and we take the action on the 1-cycles to be $[a_i]\to [a_i]$, $[b_i]\to -[b_i]$ (although other tilted orientifold actions are possible, see appendix \ref{ap:tilted-orientifolds}). The basis of even and odd 3-cycles are
\beqa
&[\alpha_0]\, =\, [a_1][a_2][a_3] \quad , \quad & [\beta_0]\, =\, [b_1][b_2][b_3]\nonumber \\
&[\alpha_1]\, =\,[a_1][b_2][b_3] \quad , \quad & [\beta_1]\, =\, [b_1][a_2][a_3]\nonumber \\
&[\alpha_2]\, =\,[b_1][a_2][b_3] \quad , \quad & [\beta_2]\, =\, [a_1][b_2][a_3]\nonumber \\
&[\alpha_3]\, =\,[b_1][b_2][a_3] \quad , \quad & [\beta_3]\, =\, [a_1][a_2][b_3]
\eeqa
The coefficients $s_{A}^{k}$ are thus
\beqa
s_{A}^{0}\, =\, m_A^1m_A^2m_A^3\;\; ,\;\; 
s_{A}^{1}\, =\, m_A^1n_A^2n_A^3\; ,\; \;
s_{A}^{2}\, =\, n_A^1m_A^2n_A^3\; ,\; \;
s_{A}^{3}\, =\, n_A^1n_A^2m_A^3\,,
\label{coupling-coeff}
\eeqa
where $(n^i, m^i)$ denote the wrapping numbers on the $i$-th torus with coordinates $(x^i, y^i)$.

\section{Discrete gauge symmetries and SM brane constructions}
\label{sec:discrete-sm}

We now turn to the study  of discrete gauge symmetries in brane constructions of
phenomenological interest. We first review the classification of discrete gauge symmetries of the MSSM in \cite{Ibanez:1991pr}, and later study its implementation in various proposed D-brane realizations of MSSM-like models.

\subsection{Discrete gauge symmetries in the MSSM}
\label{ir-symmetries}

In \cite{Ibanez:1991pr} the possible ${\IZ_N}$ generation independent discrete symmetries of the MSSM were classified in terms of the  three generators
$R,L,A$ given  in table \ref{generators}.
Here $(Q,U,D,L,E,N,$  $H_u,H_d)$ are the MSSM quark, lepton and Higgs superfields in standard notation.
Defining
\beq
R_N\ =\ e^{i\,2\pi R/N} \ ,\ 
L_N\ =\ e^{i\,2\pi L/N} \  ,\ 
A_N\ =\ e^{i\, 2\pi A/N} \ ,
\eeq
a $\IZ_N$ gauge symmetry generator may be written as
\beq
g_N\ =\ R_N^m\times A_N^n\times L_N^p \ \ ,\ \  m,n,p=0,1,..,N-1 \ .
\eeq
This is the most general $\IZ_N$ symmetry allowing for the presence of all
standard Yukawas $QUH_u,QDH_d,LEH_d$ (and also $LH_uN_R$ in the presence of
right-handed neutrinos). Note that one can obtain further but
equivalent discrete symmetries by multiplying by some power of a discrete 
subgroup of the hypercharge generator $e^{i\,2\pi (6Y)/N}$, where we use $6Y$ to make hypercharges integer.
\begin{table}
\begin{center}
\begin{tabular}{|c|c|c|c|c|c|c|c|c|}
\hline
& $Q$ & $U$ & $D$ & $L$ & E& $N_R$  &$H_u$ & $H_d$ \\
\hline\hline
$R$ & 
0 &  -1 &  1 & 0 & 1 & -1 & 1 & -1 \\
$L$ & 
0 &  0 &  0 & -1 & 1 &1 &  0 & 0 \\
$A$ & 
0 &  0 &  -1 & -1 & 0 & 1&  0 & 1 \\
\hline
\end{tabular}
\caption{\small Generation independent generators of discrete $\IZ_N$ gauge symmetries in the MSSM.
}
\label{generators}
\end{center}
\end{table}
As discussed in \cite{Ibanez:1991hv,Ibanez:1991pr}, the mixed ${\IZ_N}\times SU(3)^2$,
${\IZ_N}\times SU(2)^2$ and mixed gravitational  anomaly  constraints yield
\beqa
nN_g\ & = & 0 \ \ ,\ \ {\rm mod}\ N \\ 
(n+p)N_g\ -\ nN_D\ &=& 0 \ \ ,\ \ {\rm mod}\ N \\
-N_g(5n+p-m) \ +\ 2nN_D \ &=&\ \eta\, \frac{N}{2} \ \ ,\ \ {\rm mod}\ N
\label{disanomal}
\eeqa
where $N_g$, $N_D$ are the number of generations and Higgs sets respectively
and $\eta=0,1$ for $N=$odd, even \footnote{In the presence of $N_g$ right-handed neutrinos, which is the
generic case in brane models,  the mixed
gravitational anomaly gets simplified to $-4nN_g+2nN_D=(\eta/2)N$ mod $N$.}.

As discussed in the introduction, only discrete {\em gauge} symmetries are expected to exist in consistent theories including gravity. Therefore, it is a relevant question to assess the conditions for the above symmetries to be discrete gauge symmetries. A necessary condition is anomaly cancellation. The  $R_2$
symmetry corresponds to the usual R-parity and it is anomaly free (in fact all $R_N$ are
anomaly free for any $N$ in the presence of right-handed neutrinos). In addition, for the $N_g=3$ physical case, there are three anomaly free ${\IZ_3}$'s: $L_3$, $R_3L_3$ and $R_3L_3^2$,
as the reader may easily check using (\ref{disanomal}). The symmetry
$B_3=R_3L_3$ was introduced in \cite{Ibanez:1991pr}  and is usually called {\it baryon triality}; it allows for dimension 4 operators violating lepton number, but not  violating baryon number, so the proton is sufficiently stable. There are also additional ${\IZ_9}$  and ${\IZ_{18}}$ anomaly free discrete symmetries \cite{Dreiner:2005rd}
which involve the $A_N$ generators.
However, imposing also the purely Abelian cubic condition of \cite{Ibanez:1991hv} and 
absence of massive fractionally charged states singles out R-parity $R_2$ and
baryon triality $B_3$.

The phenomenologically interesting couplings allowed or forbidden by these discrete symmetries are displayed in table
\ref{acoplos}. The $\IZ_6$ obtained by multiplying  $R_2$ and $B_3$ is usually called {\it hexality}  \cite{Dreiner:2005rd}
and forbids all dangerous couplings but allows for a $\mu$-term and the Weinberg operator
$LLH_uH_u$ (and hence left-handed and right-handed neutrino Majorana masses).
\begin{table}
\begin{center}
\begin{tabular}{|c|c|c|c|c|c|c|c|c|}
\hline
& $H_uH_d$ & $UDD$ & $QDL$ & $LLE$ & $LH_u$ & $LLH_uH_u$ &  $QQQL$ & $UUDE$ \\
\hline\hline
$R_2$ & 
  &   x  &  x & x  & x  &   &   &  \\
$B_3=R_3L_3$ & 
  &   x  &   &   &   &  &  x & x \\
$L_3$ & 
  &    &   x &  x & x  & x &  x & x \\
  $R_3L_3^2$ & 
    &  x   &  x  & x  &  x &  &  x & x \\
    $R_2\times R_3L_3$ & 
  &   x  &  x & x  & x  &   &  x & x \\
\hline
\end{tabular}
\caption{\small  Operators forbidden by the anomaly-free ${\IZ_2}$ and ${\IZ_3}$ symmetries.}
\label{acoplos}
\end{center}
\end{table}
This ends our review of anomaly free discrete $\IZ_N$ gauge symmetries in the MSSM.

\subsection{Discrete gauge symmetries in SM-like brane models}
\label{sm-brane-symmetries}

We turn now to the appearance of discrete gauge symmetries in  explicit SM-like brane models. As in section \ref{general-analysis}, in our examples we will concentrate in toroidal type IIA orientifolds (or orbifolds thereof)  with intersecting D6-branes, although from the context it transpires that much of the analysis holds in  more general orientifolds; for instance, in the large class of Gepner MSSM-like orientifold models constructed in \cite{Dijkstra:2004cc,Dijkstra:2004ym,Anastasopoulos:2006da}. Similar results also hold in other MSSM-like constructions as well, like type IIB orientifolds with magnetized D-branes, related to IIA models by mirror symmetry (T-duality in the toroidal setup), or in heterotic compactifications with $U(1)$ bundles 
\cite{Blumenhagen:2006ux,Blumenhagen:2006wj}. Similar analysis can in principle be carried out in other setups, like D3/D7-branes at singularities, although the presence of extra multiplets beyond the MSSM ones in these models makes the analysis more model-dependent.

Much of the analysis of $U(1)$ symmetries of (MS)SM-like orientifolds can be characterized in terms of `protomodels', i.e. the gauge groups on the relevant sets of D6-branes, and the intersection numbers pattern required to reproduce the chiral matter content. These structures can subsequently be implemented in different compactifications, based on geometric spaces (toroidal or not), or non-geometric CFT setups. Results based on the protomodel structure are largely independent on their specific realization. We first consider the implementation of MSSM discrete gauge symmetries in the different MSSM-like brane protomodels, and later turn to their realization in concrete examples, for simplicity based on toroidal orientifolds. Some of these realizations are actually non-supersymmetric, but provide a good testing ground of the implementation of diverse discrete gauge symmetries.

There are two large classes of SM-like orientifolds (toroidal or not),  depending on whether the electroweak $SU(2)_L$ group is realized from a $Sp(2)$ group or from a $U(2)$. We analyze both cases in turn.

\subsubsection{The $Sp(2)$ class}
\label{sp-class}

In this class of models there are four stacks of D-branes, denoted $a$ ({\it baryonic}), $b$ ({\it left}), $c$ ({\it right}) and $d$ ({\it leptonic}). They have $N_a=3$, $N_b=N_c=N_d=1$, but the stack $b$ is taken coincident with its orientifold image, so that the initial gauge group is $U(3)_a\times Sp(2)_b\times U(1)_c\times U(1)_d$. The chiral fermion content reproduces the SM quarks and leptons if the D6-brane intersection numbers are given by\footnote{Here and in what follows, we also use the notation $A^*$ for the orientifold image of the branes $A$.}
\beqa
I_{ab}=I_{ab^*}= \ 3 \ \ & ; &\ \ I_{ac}=I_{ac^*}= \ -3 \nonumber \\
I_{db}=I_{db^*}= \ -3 \ \ & ;  & I_{cd}=\ -3 \ ;\ I_{cd^*}=\ 3 
\label{intersmsp2}
\eeqa
with the remaining intersections vanishing. As usual, negative intersection 
numbers denote positive multiplicities of the conjugate representation.
The spectrum of chiral fermions is shown in table \ref{tabpssm}, and corresponds to three SM quark-lepton generations. In addition there are 
three right-handed neutrinos $N_R$, whose presence is generic in this kind
of constructions. At the intersections there are also complex scalars with 
the same charges as the chiral fermions \cite{Ibanez:2001nd}; in supersymmetric realizations, some of these scalars are massless and complete the matter chiral multiplets, while in non-supersymmetric realizations they are generically massive (their possible tachyonic character can be avoided by a judicious choice of the complex structure moduli in concrete examples, see \cite{Ibanez:2001nd} for the toroidal case).

%
\begin{table}[htb] \footnotesize
\renewcommand{\arraystretch}{1.25}
\begin{center}
\begin{tabular}{|c|c|c|c|c|c|c|}
\hline Intersection &
 Matter fields  &   &  $Q_a$  &   $Q_c $ & $Q_d$  & Y 
\\
\hline\hline (ab),(ab*) & $Q_L$ &  $3(3,2)$ & 1  & 0 & 0 & 1/6 \\
\hline (ac) & $U_R$   &  $3( {\bar 3},1)$ &  -1   & 1  & 0 & -2/3
\\
\hline (ac*) & $D_R$   &  $3( {\bar 3},1)$ &  -1    & -1  & 0 & 1/3
\\
\hline (bd),(b*d) & $ L$    &  $3(1,2)$ &  0    & 0  & -1 & -1/2 \\
\hline (cd) & $E_R$   &  $3(1,1)$ &  0   & -1  & 1  & 1 \\
\hline (cd*) & $N_R$   &  $3(1,1)$ &  0   & 1  & 1  & 0  \\
\hline \end{tabular}
\end{center} \caption{\small Standard model spectrum and $U(1)$ charges in the 
realization in terms of D6-branes with intersection numbers (\ref{intersmsp2}).
\label{tabpssm} }
\end{table}
%

One linear combination of the three $U(1)$'s, i.e. 
\beq
Y \ =\ \frac {1}{6}\left( Q_a\ -\ 3Q_c\ +\ 3Q_d\right) \, ,
\label{hyperimr}
\eeq
corresponds to the hypercharge generator; it is anomaly free, and should be required to be massless, namely its $BF$ couplings should vanish. In the language of section \ref{general-analysis}, we have
\beqa
s_{a}^{k}\, -\,s_{c}^{k}\, +s_{d}^{k}\, =\, 0\quad \mbox{for all $k$}.
\label{hyper-massless}
\eeqa
where we have accounted for a factor of $N_a=3$ in the $s_{a}^{k}$ term, and have recalled that $N_c=N_d=1$.
Another one, $(3Q_a-Q_d)$ is anomalous (with anomaly canceled by the Green-Schwarz mechanism) and 
becomes massive as usual. The remaining orthogonal linear combination $Y'$ 
is anomaly free and will become massive or not depending on the structure 
of the couplings of the $U(1)$'s to the RR 2-forms in the given model.
Note that one can identify the generators of the previous section 
as $R=-Q_c$, $L=Q_d$ and $Q_a=3B$, with $B$ the baryon number.
There is no analogue of the $A$ generator in this class of models due to the
absence of a $U(1)_b$ associated to the electroweak group. 

Depending on the structure of the $B\wedge F$ couplings in the model, it is possible to realize the following discrete symmetries: 

\begin{itemize}

\item {\bf$R_N$ symmetries}

Since $R=-Q_c$, a $R_N$ symmetry will appear if $s_{c}^{k}\in N\IZ$ for all $k$ in the model. In particular standard R-parity will appear if $s_{c}^{k}\in 2\IZ$ for all $k$.

\item {\bf $L_N$ symmetries}

Again, since $L=Q_d$ in the brane notation, a $L_N$ symmetry will appear if $s_{d}^{k}\in N\IZ$ for all $k$ in the model. 

\item {\bf Baryon triality}

One can study the realization of combinations like $B_3=R_3L_3$. Using the above results, this requires the condition $s_{c}^{k}+s_{d}^{k}\in 3\IZ$, for all $k$. Now from (\ref{hyper-massless}) this is equivalent to the condition $s_{a}^{k}\in3\IZ$ for all $k$. An equivalent derivation is that $B_3$ can be related to baryon number $B$ by
\beqa
B_3=2Y/3-B/3
\eeqa
In any SM-like D-brane model, baryon number is realized as $U(1)_a$, and hence $B_3$ arises from its $\IZ_9$ subgroup. Due to the additional multiplicity of $N_a= 3$, this only requires $s_{a}^{k}\in 3\IZ$ for all $k$ in the model.

\item Other combinations may be studied analogously.

\end{itemize}

Let us now illustrate this in specific examples. Consider the class of non-SUSY SM-like models constructed in \cite{Ibanez:2006da}, based on a toroidal orientifold of the kind described in section \ref{toroidal-general}. Consider a set of SM branes with wrapping numbers as shown in table \ref{newfamily}. 
%
\begin{table}[htb] \footnotesize
\renewcommand{\arraystretch}{2.0}
\begin{center}
\begin{tabular}{|c||c|c|c|}
\hline
 $N_i$    &  $(n^1,m^1)$  &  $(n^2,m^2)$   & $(n^3,m^3)$ \\
\hline\hline $N_a=3$ & $(1,0)$  &  $(n_a^2, 1)$ &
 $(N_g,  m_a^3)$  \\
\hline $N_b=1$ &   $(0,1)$    &  $ (1,0)$  &
$(0,-1)$   \\
\hline $N_c=1$ & $(n_c^1,1)$  &
 $(1,0)$  & $(0,1)$  \\
\hline $N_d=1$ &   $(1,0)$    &  $(n_d^2,-N_g)$  &
$(1, m_d^3)$   \\
\hline \end{tabular}
\end{center} \caption{\small D6-brane wrapping numbers giving rise to a SM spectrum.}
\label{newfamily} 
\end{table}
Here $n_a^2$, $m_a^3$, $n_c^1$, $n_d^2$, $m_d^3$ are integers. 
The brane $b$ is mapped to itself under the orientifold action, so that the corresponding
gauge group is $Sp(2)$, identified with $SU(2)_L$.
It is easy to check that indeed these wrapping numbers give rise to the 
chiral spectrum of a SM with $N_g$ quark/lepton generations, as in table \ref{tabpssm}.
The hypercharge remains massless as long as 
\beq
n_c^1\ =\ n_a^2m_a^3+n_d^2m_d^3 \ .
\label{condiy1}
\eeq
The other two linear combinations are generically massive.  RR tadpoles cancel in this  model if
\beq
3m_a^3\ =\ N_g m_d^3 \ \ .
\eeq
In addition one should add $(3n_a^2N_g+n_d^2-16)$ D6-branes (or 
antibranes, depending on the sign) along the orientifold plane. They
have no intersection with the rest of the branes and do not modify 
the discussion in any way.

In this model, the non-vanishing $BF$ couplings from (\ref{coupling-coeff}) are
\beqa
F^a &\wedge & 3(N_gB_2^2\ +\ n_a^2m_a^3B_2^3)\nonumber \\
F^c & \wedge & n_c^1B_2^3 \nonumber \\
F^d &\wedge & (-N_gB_2^2+n_d^2m_d^3B_2^3) \, ,
\eeqa
where we have denoted $B_2^p$, $p=0,1,2,3$ the RR 2-forms. It is easy to see that this structure naturally contains some of the discrete gauge symmetries discussed above:

{\it i)} Baryon triality is quite generic. Indeed, the ${\IZ_9}$ required for matter parity appears automatically for the physical case $N_g=3$ as long as $n_a^2m_a^3$ is multiple of 3.  More generally, a ${\IZ_{N_g}}$  discrete baryon symmetry
will be present if $n_a^2m_a^3$ is multiple of $N_g$.

{\it ii)} Since $R=-Q_c$, the $R_N$ discrete symmetries (including R-parity) are naturally generated with $N=n_c^1$.

{\it iii)}  Similarly, since $L=Q_d$,  a  $L_{N_g}$ discrete symmetry appears whenever $n_d^2m_d^3$ is a multiple of $N_g$.

{\it iv)} The symmetry $R_3L_3^2$ is a $\IZ_3$ subgroup of the $U(1)$ generated by
$Q_c+Q_d$, hence it is realized as a discrete gauge symmetry whenever $n_c^1+n_d^2m_d^3=3$. This is still compatible with  (\ref{condiy1}); for instance $n_c^1=1$, $n_d^2m_d^3=2$, $n_a^2m_a^3= -1$.

Note that some of these symmetries may be  realized simultaneously, thereby generating a larger discrete gauge symmetry group. For instance, {\it hexality}, being
a product of $R_2$ and $B_3$ will appear for $n_c^1=2$ and $ n_a^2m_a^3$ a multiple of 3. These conditions are still compatible with (\ref{condiy1}).

The above class of examples is non-SUSY, still there are scalars at the intersection (not all massless) which play the
role of squarks, sleptons and Higgs scalars, so that it makes sense the study of the couplings forbidden or allowed by discrete ${\IZ_N}$ symmetries. Also, as already emphasized, it is a useful illustration of patterns which may arise in SUSY realizations in other setups richer than toroidal orientifolds.

On the other hand, there are also supersymmetric toroidal orbifold models with electroweak symmetry realized as $SU(2)_L=Sp(2)$, and reproducing an MSSM-like matter content. Consider the MSSM-like models in \cite{Cremades:2003qj}, realized in an orientifold of $\IT^6/(\IZ_2\times \IZ_2)$ as in \cite{Marchesano:2004yq}. 
The wrapping numbers $(n_\alpha^i,m_\alpha^i)$ of the different MSSM D6$_\alpha$-branes on the different 2-tori are shown in table 
\ref{table:guay-wrappings} (ignoring the additional branes required for RR tadpole cancellation), and the resulting spectrum and charge assignments are shown in table
 \ref{table:guay-spectrum}. This corresponds to the intersection numbers  (\ref{intersmsp2}) with a trivial relabeling $d\leftrightarrow d^*$. Note that the $\IZ_2\times  \IZ_2$ orbifold truncates the gauge group on $2N_A$ D6$_A$-branes to $U(N_A)$
\begin{table}
\begin{center}
\begin{tabular}{|c||c|c|c|}
\hline
 $N_\a$  &  $(n^{1},m^{1})$  &  $(n^{2},m^{2})$
&  $(n^{3},m^{3})$ \\
\hline\hline $N_a = 6$ & $(1,0)$ & $(N_g,1)$ & $(N_g,-1)$  \\
\hline $N_b=2$ & $(0,1)$ &  $ (1,0)$  & $(0,-1)$ \\
\hline $N_c=2$ & $(0,1)$ &  $(0,-1)$  & $(1,0)$  \\
\hline $N_d = 2$ & $(1,0)$ & $(N_g,1)$ & $(N_g,-1)$  \\
\hline 
\end{tabular}
\caption{\small D-brane wrapping numbers giving rise to an $SU(3) \times SU(2) \times SU(2) \times U(1)_{B-L}$ extension of the MSSM with $N_g$ quark-lepton generations. The $\IZ_2\times \IZ_2$ orbifold truncates the gauge group on $2N_A$ D6$_A$-branes to $U(N_A)$}
\label{table:guay-wrappings}
\end{center}
\end{table}

\begin{table}
\begin{center}
\begin{tabular}{|c|c|c|c|c|c|}
\hline Sector &
 Matter fields  & $SU(3) \ti SU(2)_L \ti SU(2)_R $  &  $Q_a$  & $Q_d$  & $Q_{B-L}$ \\
\hline\hline (ab) & $Q_L$ &  $3(3,2,1)$ & 1   & 0 & 1/3 \\
\hline (ac) & $Q_R$   &  $3( {\bar 3},1,2)$ & -1  &  0  & -1/3 \\
\hline (db) & $L_L$    &  $3(1,2,1)$ &  0 & -1  & -1  \\
\hline (dc) & $L_R$   &  $3(1,1,2)$ &  0  & 1  &  1  \\
\hline (bc) & $H$   &  $(1,2,2)$ &  0  & 0  &  0  \\
\hline 
\end{tabular}
\caption{\small Left-Right MSSM spectrum and $U(1)$ charges obtained from table \ref{table:guay-wrappings}, for the particular choice $N_g=3$. The $B-L$ generator is defined as $Q_{B-L} = \frac 13 Q_a + Q_d$.}
\label{table:guay-spectrum}
\end{center}
\end{table}

It is easy to check, using (\ref{coupling-coeff}), that the $BF$ couplings  are
\beqa
F^a \  &\wedge & \ 3N_g \, (\, B_2^2\, -\, B_2^3\,) \nonumber\\
F^d \  &\wedge & \ N_g \, (\, B_2^2\, -\, B_2^3\,) \, .
\eeqa
In this model $U(1)_{B-L}$ remains as a continuous gauge symmetry, generated by $Q_a/3+Q_d$. Using a hypercharge shift, this implies that $Q_c$ has no $BF$ couplings. Hence it does not make much sense to discuss discrete $R_N$ symmetries which are contained in a continuous symmetry. 

On the other hand the realization of $B_3$ as a discrete gauge symmetry is automatic for the physical case with $N_g=3$. 
On top of a nice and simple realization of 
baryon triality in an explicit MSSM-like D-brane model, this example shows an interesting  link between this symmetry and the number of generations.
Note that alternatively, since $U(1)_{B-L}$ is a gauge symmetry of the massless spectrum, $R_3L_3^2$ also remains as a discrete symmetry.

\subsubsection{The $U(2)$ class}
\label{u2-class}

In this type of models the electroweak gauge group $SU(2)_L$ is contained in a $U(2)_b$ factor. We have again an analogous structure with
branes $a$, $b$, $c$, $d$ and a gauge group $U(3)_a\times U(2)_b\times U(1)_c\times U(1)_d$. The main difference with respect to the earlier $Sp(2)$ class is that now there is an extra $U(1)_b$ gauge boson; this continuous symmetry is anomalous, but could  in principle lead to new anomaly-free discrete ${\IZ_N}$ symmetries. Also, the assignments of the $Q_b$ charge are not family independent. This follows from the structure of intersection numbers required to reproduce the (MS)SM  matter content, 
\beqa
I_{ab}\ =\ 1 \ ,\ I_{ab^*}= \ 2 \ \ & ; &\ \ I_{ac}\ =\ -3 \ ,\ I_{ac^*}= \ -3 \nonumber \\
I_{bd}\ =\ 0(-1) \ ,\ I_{bd^*}= \ -3(2) \ \ & ; &\ \ I_{cd}\ =\ -3(3) \ ,\ I_{cd^*}= \ 3(-3)\quad
\label{intersm}
\eeqa
where two options for the intersection numbers related to leptons are considered corresponding to the examples below.
The matter content and $U(1)$ charges for the first option are shown in table \ref{IMR}, while those for the second are shown in table \ref{espectrosm2}. This classes of models have no generalization to arbitrary  numbers of generations $N_g$, since the latter is related to the number of colors by anomaly cancellation \cite{Ibanez:2001nd}.

\begin{table}[htb] \footnotesize
\renewcommand{\arraystretch}{1.25}
\begin{center}
\begin{tabular}{|c|c|c|c|c|c|c|c|}
\hline Intersection &
 Matter fields  &   &  $Q_a$  & $Q_b $ & $Q_c $ & $Q_d$  & Y \\
\hline\hline (ab) & $Q_L$ &  $(3,2)$ & 1  & -1 & 0 & 0 & 1/6 \\
\hline (ab*) & $q_L$   &  $2( 3,2)$ &  1  & 1  & 0  & 0  & 1/6 \\
\hline (ac) & $U_R$   &  $3( {\bar 3},1)$ &  -1  & 0  & 1  & 0 & -2/3 \\
\hline (ac*) & $D_R$   &  $3( {\bar 3},1)$ &  -1  & 0  & -1  & 0 & 1/3 \\
\hline (bd*) & $ L$    &  $3(1,2)$ &  0   & -1   & 0  & -1 & -1/2  \\
\hline (cd) & $E_R$   &  $3(1,1)$ &  0  & 0  & -1  & 1  & 1   \\
\hline (cd*) & $N_R$   &  $3(1,1)$ &  0  & 0  & 1  & 1  & 0 \\
\hline \end{tabular}
\end{center} 
\caption{\small Standard model spectrum and $U(1)$ charges corresponding to the first choice of intersection numbers in (\ref{intersm}).}
\label{IMR} 
\end{table}

The identification of discrete symmetries $R$, $L$ is similar to section \ref{sp-class}. The symmetry $R_N$ is associated to the generator $-Q_c$, while $L_N$ is associated to the generator $Q_d$. On the other hand, the presence of $U(1)_b$ allows the realization of an axial symmetry, given by  a generation-dependent version of the $A_N$ symmetry in section \ref{ir-symmetries}. Consider for concreteness the charge assignments in table \ref{IMR}. Higgs scalars appear at $(bc)$, $(bc^*)$ intersections if branes $b$, $c$ overlap in the second torus. They have  $(Q_b,Q_c)$ charges $\pm(1,-1)$ if they arise from $(bc)$ intersections and $\pm(1,1)$ if they come from $(bc^*)$ intersections. In any event, there are Yukawa couplings for some of the quark families which are forbidden by $U(1)_b$ (whose charges, as mentioned, are generation dependent). Since the $A_N$ symmetry in section \ref{ir-symmetries} was constructed to preserve the Yukawa couplings, we try to realize a 
 discrete symmetry, which correspond to $A_N$ for those families with allowed Yukawa couplings. For instance, taking the latter to correspond to the single  generation  from $I_{ab}=1$, the $A_N$ charge assignments are reproduced by the symmetry
\beq
{\tilde A}\ =\ \frac{1}{2}(Q_a+Q_b+Q_c+Q_d) \ .
\label{axialesIMR}
\eeq
Note that this ${\tilde A}$ generator does not appear in the class of $Sp(2)$ models
we considered before. 

A clarification is in order here. The above linear combination has non-integer coefficients, contrary to our normalization (\ref{lincomb}). Actually this follows because any SM field arises from a string with both endpoints on the branes $a$, $b$, $c$ or $d$, so its charge under $Q_a+Q_b+Q_c+Q_d$ is even. The factor of $1/2$ in (\ref{axialesIMR}) brings back the normalization to minimum unit charge. Note however that other possible (potentially massive) states in the full theory, arising from strings stretching between the SM and hidden branes, would have fractional charge assignments under ${\tilde A}$. Namely, taking into account all fields in the string model, 
we should normalize the combination as $Q_a+Q_b+Q_c+Q_d$,  according to (\ref{lincomb}).  However, a $\IZ_{2N}$ subgroup acts only as a $\IZ_N$ symmetry in the SM fields, identified with the generator (\ref{axialesIMR}).

\medskip

In order to study the appearance of diverse discrete gauge symmetries, we turn to concrete explicit realizations of the above protomodels, in the toroidal setup for simplicity.
A large number of three generation toroidal non-SUSY SM-like models with intersection numbers realizing the first option in (\ref{intersm}) were constructed in \cite{Ibanez:2001nd}. The wrapping numbers of the SM D6-branes in this family of models 
are given in table \ref{solution}. The models are parametrized by a phase $\epsilon =\pm1$, four integers $n_a^2,n_b^1,n_c^1,n_d^2$ and a parameter $\rho=1,1/3$. In addition, $\beta^i=1,1/2$ depending on whether the corresponding tori are
tilted or not; the third torus is tilted for the whole class. The massless chiral spectrum is shown in table \ref{IMR}.
%
\begin{table}[htb] \footnotesize
\renewcommand{\arraystretch}{1.5}
\begin{center}
\begin{tabular}{|c||c|c|c|}
\hline
 $N_i$    &  $(n_A^1,m_A^1)$  &  $(n_A^2,m_A^2)$   & $(n_A^3,m_A^3)$ \\
\hline\hline $N_a=3$ & $(1/\beta ^1,0)$  &  $(n_a^2,\epsilon \beta^2)$ &
 $(1/\rho ,  1/2)$  \\
\hline $N_b=2$ &   $(n_b^1,-\epsilon \beta^1)$    &  $ (1/\beta^2,0)$  &
$(1,3\rho /2)$   \\
\hline $N_c=1$ & $(n_c^1,3\rho \epsilon \beta^1)$  &
 $(1/\beta^2,0)$  & $(0,1)$  \\
\hline $N_d=1$ &   $(1/\beta^1,0)$    &  $(n_d^2,-\beta^2\epsilon/\rho )$  &
$(1, 3\rho /2)$   \\
\hline \end{tabular}
\end{center} 
\caption{\small D6-brane wrapping numbers giving rise to a SM spectrum through the  first choice of intersection numbers in (\ref{intersm}), as in \cite{Ibanez:2001nd}.
\label{solution} }
\end{table}
%

Since there are tilted tori, the computation of the conditions for discrete gauge symmetries requires the results from  appendix \ref{ap:tilted-orientifolds} (note that in table \ref{solution} the labels $m_A^i$ for tilted tori actually denote the corresponding tilded quantities of appendix  \ref{ap:tilted-orientifolds}). These models have in principle up to four $U(1)$ gauge fields, but generically three of them acquire St\"uckelberg masses  due to the $B\wedge F$ couplings. The hypercharge generation is given by the same linear combination (\ref{hyperimr}), and its masslessness requires the condition
\beq
n_c^1\ =\ \frac {\beta^2}{2\beta^1}(n_a^2\ +\ 3\rho n_d^2)
\label{imrYmassless}
\eeq
Two of the three remaining $U(1)$'s are anomalous and massive, and the
third one is anomaly free and generically massive, although it may become massless
for some choices of wrapping numbers. The relevant $B\wedge F$ couplings are
\beqa
F^a &\wedge & 3\, \left(\, \frac {1}{\rho }\ B_2^2\ +\  {n_a^2} \frac {B_2^3}{2}\,\right) \nonumber\\
F^b & \wedge & 2\left(\, - \ B_2^1\ +\   {3\rho n_b^1} \frac {B_2^3}{2}\,\right) \nonumber \\
F^c & \wedge &  2n_c^1 \  \frac {B_2^3}{2} \\
F^d &\wedge & \left(\,- \frac {1}{\rho }\ B_2^2\ +\ 3\rho n_d^2\ \frac {B_2^3}{2}\,\right) \nonumber 
\eeqa
where we have taken $\beta^1=\beta^2=\epsilon=1$ to simplify the expressions, since no new interesting possibilities appear by relaxing those conditions. Note that the factor $1/2$ multiplying $B_2^3$ arises because of the tilting of the third torus; on the other hand, this tilting simultaneously leads to a factor of 2 in the actual shift of the RR scalar dual $a_3$, as compared with the coefficient of the $F_AB_2^3$ coupling.

The set of discrete gauge symmetries in this case is quite analogous to the previous $Sp(2)$, although now the symmetries cannot be generalized beyond $N_g=3$:

{\it i)} Baryon triality is obtained for $\rho=1/3$ if in addition $n_a^2$ is multiple of 3.

{\it ii)} $R_N$ discrete symmetries  with $N$ even are naturally generated with $N=2n_c^1$. In particular, R-parity is automatically implemented in all models in this class.

{\it iii)}   The $L_{3}$ discrete symmetry appears whenever $\rho=1/3$ and $n_d^2$ is a multiple of three.

{\it iv)} Note that the combination $U(1)_{\tilde A}$ in (\ref{axialesIMR}), including the factor $1/2$, has  coupling \mbox{$F_{\tilde A}\wedge (-B_2^1+\ldots)$}. This means that  there is no discrete  gauge ${\tilde A}_N$ symmetry that can be realized. This is in fact expected, since such symmetries are anomalous for $N<9$, as already mentioned. Still, it might be possible that such symmetries participate in some anomaly free combination, although we have not found any in a preliminary  search.

Note that there is a seemingly new ${\IZ_2}$ symmetry coming from $U(1)_b$. However,  it is just the center of the $SU(2)_L$ group, and as already discussed in section \ref{bf-general}, does not lead to any useful new discrete gauge symmetry.
 
Again hexality arises if $n_c^1=1$, $\rho=1/3$ and $n_a^2$ is multiple of three. These conditions are still consistent with (\ref{imrYmassless}).

\medskip

There are also fully supersymmetric models with $SU(2)_L$ as a subgroup of a $U(2)$ realizing the second option for the intersection numbers in (\ref{intersm}). 
Consider the MSSM-like models in \cite{Cremades:2002cs}, realized in an orientifold of $\IT^6/(\IZ_2\times \IZ_2)$  in \cite{Font:2006na,Ibanez:2008my}. 
The wrapping numbers are shown in table \ref{SUSYmodel2} and the massless spectrum and $U(1)$ charges 
in table \ref{espectrosm2}. It is easy to find additional branes so that all RR-tadpoles cancel \cite{Font:2006na}.
Note that the $\IZ_2\times \IZ_2$ orbifold truncates the gauge group on $2N_A$ D6$_A$-branes to $U(N_A)$

 \begin{table}[htb] \footnotesize
\renewcommand{\arraystretch}{1.25}
\begin{center}
 \begin{tabular}{|c|c|c|c|c|c|c|c|} 
\hline Intersection & 
 Matter fields  &   &  $Q_a$  & $Q_b $ & $Q_c $ & $Q_d$  & $Q_Y$ \\ 
\hline 
\hline $ab$ & $Q_L$ &  $(3, 2)$ & 1  & -1 & 0 & 0 & 1/6 \\ 
\hline $ab^*$  & $q_L$ & $2(3,2)$ &  1  & 1  & 0  & 0  & 1/6 \\ 
\hline $ac$ & $U_R$ &  $3( {\bar 3},1)$ &  -1  & 0  & 1  & 0 & -2/3 \\ 
\hline $ac^*$  & $D_R$   &  $3({\bar 3},1)$ &  -1  & 0  & -1  & 0 & 1/3 \\ 
\hline $bd$ & $ L $ &  $(1,2)$ &  0   & -1 & 0  & 1 & -1/2 \\ 
\hline $bd^*$ & $ l $ &  $2(1,2)$ &  0  & 1   & 0  & 1 & -1/2 \\ 
\hline $cd$ & $N_R$ & $3(1,1)$ &  0  & 0  & 1  & -1  & 0   \\ 
\hline $cd^*$ & $E_R$  & $3(1,1)$ &  0  & 0  & -1  & -1  & 1 \\ 
\hline $bc$ &  $H_d$ & $(1,2)$ &  0 & -1 & 1  &  0 & -1/2 \\ 
\hline $bc^*$ & $ H_u$ &  $(1,2)$ & 0 & -1 & -1 & 0  
& 1/2 \\ 
\hline 
\end{tabular} 
\end{center}
\caption{\small Chiral spectrum of the SUSY SM's of the $U(2)$ class, arising from
the second choice of intersection numbers in (\ref{intersm}).}
\label{espectrosm2}
\end{table}

\begin{table}[htb] \footnotesize
\renewcommand{\arraystretch}{1.25}
\begin{center}
\begin{tabular}{|c||c|c|c|} 
\hline  
 $N_i$    &  $(n_i^1,m_i^1)$  &  $(n_i^2,m_i^2)$   & $(n_i^3,m_i^3)$ \\ 
\hline\hline   
 $N_a=6$ & $(1,0)$  &  $(3,1)$ &  $(3 ,  -1/2)$  \\ 
\hline  $N_b= 4$ &   $(1,1)$   &  $ (1 ,0)$ & $(1,-1/2)$   \\ 
\hline  $N_c=2$ & $(0,1)$  & $(0,-1)$  & $(2,0)$  \\ 
\hline $N_d=2$ &   $(1,0)$    &  $(3,1 )$ & $(3, -1/2)$   \\ 
\hline 
\end{tabular} 
\end{center}
\caption{\small D6-brane wrapping numbers realizing the second choice of intersection numbers in (\ref{intersm}), thus leading to (a SUSY version of) the SM spectrum in table \ref{espectrosm2}. The $\IZ_2\times \IZ_2$ orbifold truncates the gauge group on $2N_A$ D6$_A$-branes to $U(N_A)$.
} 
\label{SUSYmodel2}
 \end{table}
In this example there are two massive  and two massless $U(1)$'s, including hypercharge
and $B-L$. The $B\wedge F$ couplings are
\beqa
F^a &\wedge & 9(B_2^2\ - \frac {B_2^3}{2}) \\
F^b & \wedge &  2 (B_2^1- \frac {B_2^3}{2})\\
F^d &\wedge &  3(B_2^2\ - \frac {B_2^3}{2}) \ .
\eeqa
Again, the third $\IT^2$ is tilted, so the coefficient of the $F_AB_2^3$ coupling receives an additional factor of 2 upon dualization to a shift of the dual RR scalar; this effectively removes the factors $1/2$ accompanying $B_2^3$.

Note that again in this example baryon triality $B_3$  is automatic and so is $L_3$.
Also, no new non-trivial discrete symmetries arise from the presence of a $U(1)_b$ gauge symmetry.

In summary, discrete gauge symmetries are endemic in  SM- and MSSM-like brane constructions. R-parity (and $R_N$ extensions), baryon triality $B_3$ and lepton triality  $L_3$ appear generically  in large classes of models. In the MSSM-like examples considered, baryon triality and lepton triality  appear automatically and R-parity is extended to a full continuous $U(1)_{B-L}$ group.  It is remarkable that all anomaly free ${\IZ_2,\IZ_3}$ discrete symmetries of the MSSM classified in \cite{Ibanez:1991pr} appear in brane models.
No larger anomaly free ${\IZ_9,\IZ_{18}}$  discrete symmetries \cite{Dreiner:2005rd} are generated. That might be due to the fact that those symmetries involve the
$A_N$ generators  which do not occur  in the models examined.

\subsection{Discrete gauge symmetries and $SU(5)$ unification}
\label{sec:su5}

It is interesting to explore whether discrete gauge symmetries also appear in models
with a unified gauge symmetry like $SU(5)$. It is possible to construct type II orientifolds with
a $SU(5)$ gauge group and appropriate matter and SM Higgs multiples. However, in these models the Yukawa couplings $10\times 10\times 5_H$ can only appear at the non-perturbative level, since they violate the $U(1)\subset U(5)$ symmetry, which is perturbatively exact. Such couplings are on the other hand easy to obtain in the context of F-theory GUT's, see section \ref{sec:fth}. In the framework of type II orientifolds, they can be generated by D-brane instanton effects, see \cite{Blumenhagen:2007zk,Ibanez:2008my} for further discussion. The fact that certain instantons must play an important role in this class of models gives an added interest to the question of whether certain phenomenologically undesirable operators are protected against analogous non-perturbative effects; discrete gauge symmetries are the perfect tool to enforce such property.

For instance, a potential problem of generic $SU(5)$ unification models is the presence of dimension 4  couplings $10\cdot {\bar 5}\cdot {\bar 5}$, which contain $UDD$, $DQL$ and $LLE$ couplings, giving rise to  fast proton decay. Other potentially dangerous dimension 5 couplings are $10\cdot 10\cdot 10\cdot {\bar 5}$ which contain the operators $QQQL$ and $UUDE$ which may also give rise to too fast proton decay.  We would like to see whether discrete symmetries forbidding these couplings are generated in brane models.

A large set of Gepner model $SU(5)$ orientifolds models was studied in \cite{Anastasopoulos:2006da,Kiritsis:2009sf,Anastasopoulos:2010hu}. We will restrict, however, to a study of the $U(1)$ symmetries in the simplest intersecting D-brane setting
which may contain $SU(5)$ unification as  described in e.g. \cite{Blumenhagen:2007zk}.
Consider a stack $1$ of five D6-branes with gauge group $U(5)_1$ intersecting a single D6-brane $2$ 
with gauge group $U(1)_2$. The minimal structure of D6-brane intersections required to get a $SU(5)$ GUT is as in table \ref{tablasu5D6}.
\begin{table}[htb] \footnotesize
\renewcommand{\arraystretch}{1}
\begin{center}
\begin{tabular}{|c|c|c|}
\hline 
 Intersection  &  $I_{ab}$ &  $U(5)_1\times U(1)_2$ 
\\
\hline\hline   $11^*$ &  3  &  ${\bf 10}_{(2,0)}$ \\
\hline   $12$  & 3 &  ${\bf \bar 5}_{(-1.1)}$ 
\\  
\hline
 $22^*$   &  3  &  ${\bf 1}_{(0,-2)}$ \\ 
\hline
 $12^*$ & 1 & ${\bf 5}^H_{(1,1)} + {\bf \bar 5}^H_{(-1,-1)}$ \\ 
\hline
\end{tabular}
\end{center}
\caption{\small  Configuration of  intersecting D6-branes realizing
an $SU(5)$ GUT. }
\label{tablasu5D6}
\end{table}
The subindices show the $U(1)_1\times U(1)_2$ charges, and asterisks  denote orientifold image D6-branes. The ${\bf 10}$'s and ${\bf{\bar 5}}$'s arise from the $11^*$  and $1^*2^*$ intersections, with $I_{11^*}=3$, $I_{12}=-3$, respectively, whereas the Higgs fields reside at  $12^*$ intersections. The 
D-type Yukawas
$ {\bf 10}_{(2,0)}\cdot {\bf \bar 5}_{(-1,1)} \cdot {\bf \bar 5}^H_{(-1,-1)}$ 
are allowed by the $U(1)$ symmetries, whereas the U-type coupling
${\bf 10}_{(2,0)}\cdot {\bf 10}_{(2,0)} \cdot {\bf 5}^H_{(1,1)}$
 is forbidden.

Since neither $L$ nor $B$ generators commute with $SU(5)$, it is not possible to generate symmetries like baryon triality or lepton triality as discrete symmetries of the $SU(5)$ model. However it is easy to obtain R-parity or some ${\IZ_N}$ generalization, as discrete subgroups of the generator\footnote{Concerning the coefficient $1/2$, the same comments as for ${\tilde A}$ in (\ref{axialesIMR}) apply.}
 \beqa
  Q_X\ =\ \frac {1}{2} ( Q_1\ - \ 5Q_2) \ =\  5(B-L) \ -\  4Y  \ .
 \label{qx}
 \eeqa
 This is the familiar $U(1)$ in the branching $SO(10)\rightarrow SU(5)\times U(1)$, under which the ${\bf 16}$ and the ${\bf 10}$ decompose as
 \beqa
 {\bf 16} &\; \rightarrow\; & {\bf 10}_{1}\, +\,{\bf{\ov 5}}_{-3}\, +\, {\bf 1}_{5}\nonumber \\
 {\bf 10} &\; \rightarrow\; & {\bf 5}_{2}\, +\, {\bf{\ov 5}}_{-2}
 \eeqa
 Therefore if $\frac 12(s_1^k-5s_2^k)\in N\IZ$, then the $BF$ couplings imply that a ${\IZ_N}$ subgroup of $U(1)_X$ survives as a discrete gauge symmetry. This suffices to 
forbid the $L$- and $B$-violating couplings in $10\cdot {\bar 5}\cdot {\bar 5}$.  For  $N=2$ one recovers the usual R-parity, since $Q_X$ is mod 2 equal to $B-L$ (up to hypercharge shift), whereas for e.g. $N=4$ one recovers a ${\IZ_4}$ symmetry first  suggested by Krauss and Wilczek   \cite{Krauss:1988zc}. These generalizations of R-parity have however the shortcoming of forbidding neutrino Majorana masses. 
 
 In principle there could also be discrete symmetries coming from the orthogonal $U(1)$ symmetry
 \beq
 Q_{Z} \ =\ \frac {1}{2} (\,5Q_1 \, +\,  Q_2\,)
 \eeq 
 under which the fields have charges $10_5$, ${\bar 5}_{-2}$, $1_{1}$, ${5^H_{3}}$, ${\bar 5}^H_{-3}$. A  ${\IZ_2}$ subgroup
 of $Q_Z$ would allow for neutrino Majorana masses but would forbid the instanton generation of U-quark Yukawas, since 
 all fields would be odd except for ${\bar 5}$. 
 
 Thus within this type of brane configurations R-parity (or ${\IZ_N}$ generalizations) may in principle appear as a discrete gauge symmetry. However additional discrete symmetries will typically forbid either the generation of U-quark Yukawas or neutrino Majorana masses or both. It seems also difficult to forbid 
 dim=5 couplings $10\times 10\times 10\times {\bar 5}$ without forbidding at the same time U-quark Yukawa couplings. It would be interesting to see whether these conclusions based on the simplest D-brane configuration remain true in more general cases.
 
\section{Discrete gauge symmetries in local F-theory GUTs}
\label{sec:fth}

F-theory can be regarded as a non-perturbative generalization of type IIB compactifications with D7-branes. In the same spirit, local F-theory GUTs can be regarded as a non-perturbative generalization of type IIB models with GUT theories localized on stacks of D7-branes. However, a key difference in both situation is the status of $U(1)$ symmetries (and so, for instance, the presence or not of certain couplings, like the up-type Yukawa in $SU(5)$ theories). Since $U(1)$ symmetries are so intimately linked with discrete gauge symmetries, it is worthwhile to explore the extension  to the realm of F-theory of our earlier description of discrete gauge symmetries in D-brane models. This would place important restrictions on the very active topic of brane instanton effects in F-theory (see e.g. \cite{Blumenhagen:2010ja,Cvetic:2010rq}) .

The physics of $U(1)$ gauge theories in F-theory is in general poorly understood in compact examples. We therefore focus on local F-theory GUT models, which have been extensively studied (see e.g. \cite{Donagi:2008ca,Beasley:2008dc,Beasley:2008kw,Donagi:2008kj}, and also \cite{Heckman:2010bq,Weigand:2010wm} for reviews). The starting point is provided by F-theory 7-branes wrapped on a local 4-cycle $S$ in the base of the elliptically fibered CY fourfold, leading to an $SU(5)$ GUT theory (with no overall $U(1)$ factor). There are other 7-branes on other 4-cycles $S_A$ (which are non-compact in the local description) which intersect $S$ along complex curves $\Sigma_a$ (matter curves). These intersections support charged matter, in representations of the gauge factors of both 7-branes dictated by the enhanced symmetry at the intersection locus. In particular, local enhancements to $SU(6)$ lead to fields in the $5$, ${\bar 5}$, and local enhancements to $SO(10)$ lead  to fields
  in the $10$, ${\ov{10}}$. In addition, local rank-2 enhancements at points of $S$, due to intersections of several matter curves, correspond to Yukawa couplings among the fields supported on the latter. In this local picture, the $U(1)_A$'s supported by the non-compact 7-branes on $S_A$ are global symmetries, which may or may not survive in a full fledged compactification, due to global geometrical effects. Even if they survive  these effects, and seemingly manifest as 4d $U(1)$ gauge symmetries, they may acquire St\"uckelberg masses by their $BF$ couplings. We are thus interested in determining {\em necessary} conditions (which are not sufficient due to this global sensitivity) for $\IZ_n$ subgroups of these $U(1)$'s to survive as discrete gauge symmetries of the model.

There are certainly many possibilities in F-theory model building. For concreteness we will focus on a particular class of models, in which there is a good control of the $U(1)_A$ charges of the different $SU(5)$ representations in the different curves; the basic ideas concerning the $BF$ couplings in F-theory however hold more generally. The models we focus on have an underlying $E_8$ structure globally on the 4-cycle $S$, in the sense that the pattern of matter curves and Yukawa points is determined by an unfolding of $E_8$ into $SU(5)$, according to
\beqa
E_8\; & \; \rightarrow \; & \quad SU(5)_{\rm GUT}\times SU(5)_\perp\nonumber\\
248 & \rightarrow & (24,1)\, +\, (10,5)\, +\, (5,{\ov {10}})\, +\, ({\ov {10}},{\ov 5})\, +\, ({\ov 5},10)\, +\, (1,24)
\eeqa
where $SU(5)_\perp$ is actually split to $U(1)^4$, but is useful as shorthand for the corresponding charges. For the $SU(5)_{\rm GUT}$ $5$'s and $10$'s, and singlets, these are specifically given by
\beqa
&\;SU(5)_{\rm GUT} \; &\quad \quad U(1)^4\nonumber \\
& 10 & \quad (\underline{4,-1,-1,-1,-1})\nonumber\\
&5 & \quad (\underline{3,3,-2,-2,-2})\nonumber\\
& 1 & \quad (\underline{1,-1,0,0,0})
\eeqa
where underlining means permutation of entries; also, conjugate $SU(5)_{\rm GUT}$ representations have opposite $U(1)^4$ charges. Note that we have represented charges with respect to five $U(1)$'s with generators $Q_A$ $A=1,\ldots, 5$ but constrained by $\sum_A Q_A=0$ (corresponding to Cartan generators of $SU(5)_\perp$). 

This class of models has been extensively discussed in e.g. \cite{Hayashi:2009ge,Donagi:2009ra,Marsano:2009ym,Heckman:2009mn,Marsano:2009gv,Blumenhagen:2009yv,Marsano:2009wr,Dudas:2009hu,Grimm:2009yu,Hayashi:2010zp,Grimm:2010ez,Marsano:2010ix,Dudas:2010zb,Ludeling:2011en,Dolan:2011iu}. The global $E_8$ structure throughout $S$ allows the use of the so-called spectral cover construction, to encode most of the relevant information about the local geometry around $S$ (sometimes referred to as semi-local model), including the 7-brane worldvolume gauge fluxes. Roughly speaking, the system is a configuration of F-theory 7-branes leading to an $E_8$ gauge theory on $S$, deformed by vevs (rather, backgrounds varying along $S$) of scalars in the adjoint of $SU(5)_\perp$, thus leaving only $SU(5)$ as the unbroken group. This point-dependent $SU(5)_\perp$ matrix can be diagonalized, in terms of five (point-dependent) eigenvalues $\phi_i$ (with $\sum_i \phi_i=0$),  leading to a 5-fold co
 vering of $S$, known as spectral cover.
In general, the scalar profiles  can have poles (it is formally a meromorphic Higgs bundle), so that the extra 7-branes go off to infinity and are non-compact in the semi-local model. Also, the 5-fold cover is in general branched, meaning that some of the $U(1)$'s are related by monodromies, subgroups of $S_5$ (the group of permutations of 5 elements), that describe the reshuffling of sheets of the cover as one loops around in $S$.

This description is fleshed out by describing the semilocal geometry of the elliptic fibration  in the Tate form, describing the unfolding of $E_8$ into $SU(5)$:
\beqa
x^3\,-\,y^2\,+xyz \, b_5 w\,+\, x^2z^2\, b_4w^2 \, +\, yz^3\, b_3w^3\, +xz^4\, b_2 w^4\, +\, z^6\, b_0 w^5\, =\, 0
\label{tate}
\eeqa
where $[z,x,y]$ are homogeneous coordinates in ${\IP}_{[1,2,3]}$, parametrizing the elliptic fiber, $w$ is a coordinate transverse to $S$, and $b_i$ are functions (actually, sections of suitable line bundles) over $S$. 

The 4-cycle $S$ corresponds to the locus $w=0$, where the above equation can be shown to describe a degeneration of the elliptic fiber leading to an $SU(5)$ gauge symmetry. The information on the extra 7-branes is encoded in the $b_n$, and is nicely captured by the $SU(5)$ spectral cover ${\cal C}_5$, a 5-sheeted branched cover of $S$ living in an auxiliary non-CY threefold $X$; the latter is defined as a $\IP_1$ bundle over $S$, 
$\IP({\cal O}_S\oplus K_S)$, where ${\cal O}_S$ and $K_S$  are the trivial and canonical line bundles over $S$, respectively. The spectral cover is defined by the equation
\beqa
b_0\, s^5\, +\, b_2\, s^3\, +\, b_3\, s^2\, +\, b_4\, s\,+\, b_5\,=\, 0
\label{su5-cover}
\eeqa
where $s$ is an affine coordinate in the $\IP_1$ fiber in $X$, so $S$ is defined by $s=0$. The $b_n$ are symmetric monomials in some variables $\phi_i$, regarded as the Higgs vevs, with $b_1=0$ due to tracelessness of the $SU(5)_\perp$ generators. The spectral cover ${\cal C}_5$ contains the information about the matter curves, for instance the ${\bf 10}$ matter curves are associated to its intersection with $S$; this is the locus $b_5=0$, which can be shown to correspond in (\ref{tate}) to a locus of enhanced $SO(10)$ symmetry. The ${\bf 5}$ curves arise from an associated spectral cover ${\cal C}_{10}$, describing the representation of the Higgs field in the ${\bf 10}$ of $SU(5)_\perp$. There are several techniques to compute the location of the different matter curves, and their homology classes, for which we refer the reader to the references. 

The spectral cover is particularly useful to characterize the 7-brane worldvolume $U(1)_A$ fluxes, required to obtain chiral matter from the 6d multiplets localized on the matter curves. This is done in terms of a suitable line bundle ${\cal N}_5$ over the spectral cover ${\cal C}_5$. Once projected down to $S$, this defines an $SU(5)_\perp$ bundle $V$ over $S$, which can be regarded as fully responsible for the breaking of the underlying $E_8$ symmetry to its commutant $SU(5)_{\rm GUT}$. In addition, the line bundle can include components corresponding to hypercharge flux $F_Y$, in order to break $SU(5)_{\rm GUT}$ to the SM group. As emphasized in \cite{Beasley:2008kw} (see \cite{Buican:2006sn} for an earlier realization in a different context), masslessness of the hypercharge gauge boson requires the 2-form $F_Y$ to be non-trivial on $S$, but trivial in the global geometry. Since this prevents $F_Y$ to have $BF$ couplings to bulk 2-forms, its introduction is irrelevant for 
 the purpose of studying discrete gauge symmetries, and we ignore it in the following.

To our knowledge, the computation of $BF$ couplings for the $U(1)^4$ factors in F-theory has not been carried out in detail in the literature in the spectral cover language. However, they are easily guessed to arise from  a Chern-Simons (CS) coupling on the  F-theory 7-brane worldvolume 
\beqa
\int_{S_A\times M_4}C_4\wedge F_A\wedge F_A
\label{fth-cs}
\eeqa
This can be regarded as a simple generalization of the CS couplings on D7-branes. More rigorously, it can be easily  derived from the dual picture of M-theory on a CY fourfold. The degenerations of the elliptic fiber on top of the 7$_A$-branes support harmonic 2-forms $\omega_A$, normalized to $\int_{\rm ALE} \omega_A\wedge \omega_B=\delta_{AB}$, where ALE stands for the local ALE geometry transverse to the degeneration locus. The component of the M-theory 3-form $C_3$ along $\omega_A$ becomes the $7_A$ worldvolume gauge field, so its field strength $G_4=dC_3$ has a component
\beqa
G_4\, =\, \sum_A \omega_A\wedge F_A
\label{magnet-fth}
\eeqa
The 11d effective action of M-theory has a Chern-Simons coupling
\beqa
\int_{11d} C_3\wedge G_4\wedge G_4
\eeqa
Upon replacing (\ref{magnet-fth}), and noticing that the M-theory $C_3$ maps to the F-theory $C_4$ under duality, we recover (\ref{fth-cs}).

The coupling (\ref{fth-cs}) produces the relevant $BF$ couplings for $U(1)_A$, as follows. We introduce two basis of dual 2-cycles $\{\alpha_k\}$, $\{\beta_k\}$ in $S$, with $\alpha_k\cdot\beta_l=\delta_{kl}$. As suggested by the notation, they play a role analogous to the 3-cycles in section \ref{bf-general}.
Some of these cycles may be trivial in the global geometry, so in what follows we implicitly restrict the range of $k$ to globally non-trivial classes\footnote{In other words, what counts is the class of $[F]$ in the cohomology of the threefold, rather than of the 4-cycle.}. We define the 4d 2-forms
\beqa
B_k\, =\,\int_{\beta_k} C_4
\eeqa
In addition, we expand the magnetic flux of $U(1)_A$ as
\beqa
F_A\, =\sum_k s_A^k \beta_k \quad , \quad \mbox{namely $s_A^k=\int_{\alpha_k} F_A$}
\label{f-fth-exp}
\eeqa
Reduction of the coupling (\ref{fth-cs}) leads to the $BF$ terms
\beqa
\sum_A s_A^k B_k\wedge F_A
\label{fth-bf}
\eeqa
Therefore, for a linear combination $Q=\sum_A c_A Q_A$ to leave a $\IZ_n$ discrete gauge symmetry, the necessary condition is
\beqa
\sum_A c_A \,s_A^k\, = \, 0\; \mbox{mod $n$ for all $k$ (with $\beta_k$ non-trivial in global geometry)}
\label{condi-fth}
\eeqa
This agrees with the D-brane condition below (\ref{bf-gen}) with $N_A=1$, as is the case here. Also, it corresponds to the $BF$ couplings in compactifications of the heterotic string with $U(1)$ bundles \cite{Blumenhagen:2005ga,Blumenhagen:2005pm}.

The above condition is necessary, but not sufficient, for several reasons: First, the $U(1)$ may actually be broken by global effects, as mentioned. Even semi-locally, there are in general monodromies \cite{Hayashi:2009ge}, which eliminate some of the relative $U(1)$'s (e.g. $Q_1-Q_2$ for a $\IZ_2$ monodromy). For instance, a generic spectral cover ${\cal C}_5$ (\ref{su5-cover}) is irreducible, so there are $\IZ_5$ or $S_5$ monodromies that mix all sheets in the spectral cover, and leave no $U(1)$ symmetry whatsoever (since $SU(5)_\perp$ has no overall $U(1)$ factor). In order to lead to non-trivial $U(1)$ symmetries, the spectral cover must be split, with two or more disconnected components (and in fact the split should extend even globally), as we consider in upcoming examples. Note that, even if there is such a $U(1)$ symmetry, the global geometry may contain additional 2-forms, not present in the local model, coupling to the $U(1)$ with $BF$ couplings  not satisfying the 
 condition (\ref{condi-fth}).
 
An important ingredient about discrete gauge symmetries from $BF$ couplings is their anomaly cancellation. As suggested from our discussion in section \ref{sec:anomaly}, this leans on the structure of corresponding mixed $U(1)$ anomalies, and their cancellation by a Green-Schwarz mechanism. The latter has not been worked out in the F-theory context, but we may adopt a safe attitude and focus on $U(1)$ factors which are anomaly free. For $SU(5)$ theories, there is one family-independent $U(1)$ factor, already appeared in section \ref{sec:su5}. It is the generator $Q_X$, arising in the decomposition of $SO(10)\to SU(5)_{\rm GUT}\times U(1)_X$. F-theory models where this $U(1)_X$ remains as the only remnant of the original $U(1)^4$ are based on an $S(U(4)\times U(1))$ spectral cover, rather than an $SU(5)$ one. The spectral cover factorizes in two reducible pieces ${\cal C}_4$, ${\cal C}_1$, with (\ref{su5-cover}) now having an structure
\beqa
(\, c_0 \,s^4\, c_1\, s^3\, +\,c_2\, s^2\,+\, c_3\, s\,+\, c_4\,)\,(\, d_0\, s\,+\, d_1\,)\,=\,0
\eeqa
with $b_1=c_0d_1+c_1d_0=0$. This means that four sheets of the spectral cover mix among themselves, while the last remains factorized. The construction of $S(U(4)\times U(1))$ spectral covers is a generalization of that of $SU(5)$ spectral covers, carried out in \cite{Blumenhagen:2009yv}. The introduction of the 7-brane worldvolume fluxes is carried out in terms of two line bundles ${\cal N}_4$, ${\cal N}_1$ over ${\cal C}_4$, ${\cal C}_1$, which project onto $S$ as $U(4)$ and $U(1)$ bundles $V_4$, $L$, respectively. Their first Chern classes are integer cohomology classes in $S$, and are constrained by
\beqa
c_1(V_4)\,+\, c_1(L)\,=\, 0
\eeqa
so the construction actually defines an $S(U(4)\times U(1))$ bundle, with a commutant  $SU(5)\times U(1)_X$ in $E_8$.

This defines the 4d gauge group (ignoring hypercharge flux), before accounting for the $BF$ couplings of $U(1)_X$. These are controlled by $c_1(L)$, i.e. the cohomology class of $[F_X]$, considered as a class in the global geometry (rather than just in $S$). In order to show that they can indeed lead to interesting discrete gauge symmetries, we consider two explicit examples of compact models,  in \cite{Blumenhagen:2009yv,Grimm:2009yu}, leading to 3-generation $SU(5)$ GUTs (with hypercharge flux breaking to the SM), and for which the $S(U(4)\times U(1))$ structure holds even globally. 

The global example in \cite{Blumenhagen:2009yv}, is based on a base $B_3$ obtained from the Fano threefold $\IP_4[4]$ (i.e. the subspace of $\IP_4$ defined by a homogeneous equation of degree 4) by a geometric transition introducing a $dP_7$ del Pezzo 4-cycle. The basic 2-cycle classes are $H$ and $X$ (related to the hyperplane class in $\IP_4$ and the exceptional divisor $dP_7$ itself), and the K\"ahler cone is spanned by $H$ and $H+X$. A detailed construction of the elliptic fibration, and the worldvolume fluxes, led to the construction of a 3-generation F-theory $SU(5)$ GUT (broken to the SM by suitable hypercharge flux), with an additional $U(1)$ 4d gauge symmetry. The $BF$ couplings can be derived from the Fayet-Illiopoulos terms in eq. (138) in that reference, and read
\beqa
(\,-12 B_1\, +\,8\,B_2\,)\wedge F
\eeqa
where $B_1\,=\int_H C_4$, $B_2=\int_{H-X}C_4$. There is therefore a $\IZ_4$ discrete gauge symmetry, which corresponds to the generalized R-parity in \cite{Krauss:1988zc}. The model in \cite{Grimm:2009yu} has a more involved structure, but similar qualitative features. From the FI terms in eq. (5.6) in that reference, the $BF$ couplings have a structure
\beqa
(\,6 B_1\, -\,12\,B_2\,+12B_3)\wedge F
\eeqa
so there is a $\IZ_6$ discrete gauge symmetry of the generalized R-parity type. Beyond these concrete examples, there seems to be no fundamental obstruction to realizing a genuinely $\IZ_2$ R-parity in other examples constructed using similar techniques. We hope this analysis suffices to show the appearance of discrete gauge symmetries in F-theory, and leave a more systematic understanding for future work.

\section{K-theory $Z_2$ and R-parity}
\label{k-theory}

The K-theory constraints in orientifold models force some combination of quantities to be even. Interestingly enough, these quantities arise as the coefficients of some of the $BF$ couplings in the model. Hence, in certain classes of construction, the K-theory constraints imply the existence of an anomaly-free $\IZ_2$ discrete gauge symmetry, which we denote $K_2$. We now describe the conditions for its existence and also its interplay with the massless $U(1)$ possibly present in the model.

Consider an orientifold with D6$_A$-branes on a general CY orientifold, with basis  $\{\alpha_k\}$, $\{\beta_k\}$ of even and odd cycles, and assume for simplicity that $\alpha_k\cdot \beta_l=\delta_{kl}$. The K-theory constraints in the model have the structure 
\beqa\label{Kthconstr}
\sum_A N_A c_{Ak1}\, \in\, 2\IZ
\eeqa
for all $k_1$ in a subset of the odd cycles. Notice that \eqref{Kthconstr} is not necessarily imposed to all odd cycles; for instance, in orientifolds of $\IT^6$ the K-theory constraints are
\beqa
\sum_A N_A m_A^1n_A^2 n_A^3\in 2\IZ\quad , \quad
\sum_A N_A n_A^1m_A^2 n_A^3\in 2\IZ\quad , \quad
\sum_A N_A n_A^1n_A^2 m_A^3\in 2\IZ
\eeqa
whereas there is no constraint on the combination $\sum_A N_A m_a^1m_a^2m_a^3$. So $k_1$ labels the odd cycles $\alpha_1$, $\a_2$, $\a_3$, but not  $\a_0$. 

Also, not all branes contribute, i.e. some $c_{Ak}$ may be zero; for instance, the branes $b$, $c$ in the model in table \ref{table:guay-wrappings} have no contribution to the K-theory constraints. We label with $A_1$ those branes for which $c_{A_1k_1}\neq 0$, so the K-theory constraints read
\beqa
\sum_{A_1} N_{A_1} c_{A_1k1}\, \in\, 2\IZ
\label{kth-cond-red}
\eeqa
On the other hand, the $BF$ couplings have a structure
\beqa
\sum_A\sum_k\, N_A c_{Ak}\, B_k\wedge F_A\, =\, 
\sum_A\sum_{k_1}\, N_{A_1} c_{A_1k_1}\, B_k\wedge F_A\, +\,\sum_A\sum_{k_2} N_A c_{Ak_2}\, B_k\wedge F_A
\eeqa
where $k$ runs over all odd cycles, and $k_1$, $k_2$ label those with or without an associated K-theory charge cancellation constraint. Note that for the $BF$ couplings of the former kind, there are contributions only from branes with label $A_1$, i.e. participating in (\ref{kth-cond-red}).

Now assume that in the class of models under consideration, $c_{A_1k_2}=0$, for all $A_1$, $k_2$. This may sound a strong condition, but holds even in the simplest semi-realistic intersecting D6-brane models in $\IT^6$ orientifolds, see section \ref{sm-brane-symmetries}, where $k_2$ labels only the cycle $[b_1][b_2][b_3]$; its associated 2-form has $BF$ couplings proportional to $N_Am_A^1m_A^2m_A^3$, which vanishes for all branes in such models.

Under these assumptions, the diagonal combination of the $U(1)$'s contributing to the K-theory charges
\beqa
Q_{\cal K}\, =\, \sum_{A_1} \, Q_{A_1}
\label{kiu-kei}
\eeqa
has a $BF$ coupling
\beqa
\sum_{A_1} N_{A_1} c_{A_1 k_1}\, B_{k_1}\wedge F_{\cal K}
\eeqa
The K-theory constraint (\ref{kth-cond-red}) implies the existence of a $\IZ_2$ discrete gauge symmetry $K_2$.

Many models have massless $U(1)$'s, and one must ensure that $K_2$ is not just a subgroup of these. We write the massless $U(1)$ generator as
\beqa
Q\, =\, \sum_A r_A Q_A
\eeqa
with $r_A\in \IZ$ and gcd($r_A$)=1, so that charges are integer with minimal charge one. If $r_A$=odd for all $A$, then $Q=Q_{\cal K}$ mod 2, and $K_2$ is just a subgroup of the massless $U(1)$. This may seem non-generic, but occurs e.g. in many SM-like D-brane models, where the massless hypercharge generator is typically of the form
\beqa
6Y\,=\, Q_a\, -\, 3Q_c+\, 3Q_d
\label{hypercomb}
\eeqa
Since its coefficients are odd, in models where the branes $a$, $c$, $d$ contribute to the K-theory charges, the symmetry $K_2$ is just a subgroup of hypercharge.

The $\IZ_2$ discrete gauge symmetry $K_2$ receives a natural interpretation in the mirror/T-dual picture of magnetized type I compactifications, in which the K-theory charges are  (non-BPS D7-brane charges) induced on the D9-branes by their worldvolume magnetic fluxes. The K-theory constraints require the D9-brane gauge bundle to be in $Spin(32)/\IZ_2$, where the $\IZ_2$ acts as $-1$ on vector representations, i.e. corresponds to $K_2$.

It would be tempting to exploit the symmetry $K_2$ to generate a phenomenologically relevant $\IZ_2$ symmetry in MSSM-like D-brane models, e.g. R-parity. However, there are several difficulties in the simplest implementation of this idea. In most SM-like models, c.f. the earlier sections, there are four stacks of branes $a,b,c,d$, corresponding to the baryonic, left, right, and leptonic branes, respectively. R-parity can be generated as the $\IZ_2$ subgroup of $Q_a+Q_d$ (which mod 2 is equal to $B-L=Q_a+3Q_d$). To realize this as the symmetry $K_2$, we need the branes $a$, $d$ to be the only ones contributing to the K-theory charges, $c_{ak_1}, c_{dk_1}\neq 0$. Now the requirement to have massless hypercharge (\ref{hypercomb}) implies
\beqa
c_{ak}\, -\, 3c_{ck}\, +\, 3c_{dk}\, =\, 0 \quad \mbox{for all $k$}
\eeqa
Since $a,d$ are in the range of $A_1$, they are assumed to have $c_{ak_2}=c_{dk_2}=0$, and so, for $k=k_2$, we have $c_{ck_2}=0$. If $c_{ck_1}\neq 0$, then $c$ also contributes to the K-theory charge and $K_2$ is actually the $\IZ_2$ subgroup of $Q_a+Q_c+Q_d$. As mentioned above, modulo 2 this is equal to $6Y$ and $K_2$ is just part of the hypercharge $U(1)$ symmetry. If instead $c_{ck1}=0$, then $Q_c$ has no $BF$ couplings, and masslessness of hypercharge implies masslessness of $B-L=Q_a+3Q_d$. Hence $K_2$ is indeed R-parity, but is embedded as a subgroup of $U(1)_{B-L}$.

There are several possible ways to relax the constraints on the $BF$ couplings of $c$, and possibly overcome the above problems. For instance, the hypercharge combination may involve extra `hidden' $U(1)$ generators; this however invokes symmetries beyond the visible MSSM-like sector. Also, we may relax the condition $c_{A_1k_2}=0$ to $c_{A_1k_2}\in 2\IZ$, and still have a $\IZ_2$ symmetry from (\ref{kiu-kei}); however this exploits additional even-ness requirements, beyond the genuinely K-theoretical one. We thus do not pursue these possibilities here.

\section{Final comments and conclusions }

In this paper we have studied the natural appearance of discrete gauge symmetries in
large classes of string vacua, concretely those based on D-branes in type II orientifolds, and local 7-brane systems in F-theory GUTs.  They have a number of novelties as compared with earlier studies of discrete gauge symmetries in string theory, which were mainly based on heterotic string compactifications. The main advantage of the present setup is that the discrete symmetries are manifest in the model, without resorting to the rather model dependent choices of flat direction required in the heterotic setup.
Also, in the present setup these discrete gauge symmetries are,  by construction,  anomaly free
and are respected by non-perturbative instanton effects  --- although in particular systems it may physically meaningful to use (potentially anomalous) discrete symmetries preserved by some instantons, but violated by others ---.

We have shown how semi-realistic (MS)SM  type II orientifold constructions naturally 
bring in discrete gauge symmetries which are  ${\IZ_N}$ subgroups of continuous $U(1)$ symmetries in the
models.  Specifically, they correspond to discrete subgroups of baryon and lepton number $U(1)$ symmetries
(modulo discrete hypercharge rotations). The list of discrete symmetries arising is very limited and corresponds to
the anomaly free classification of discrete gauge symmetries in \cite{Ibanez:1991pr}:

{\bf i)} The discrete groups  $R_N$ which may be understood as discrete subgroups of $U(1)_{B-L}$, with $R_2$
corresponding to R-parity. These symmetries forbid all dimension four B- and L-violating couplings, although they
do not forbid the unwanted (but less dangerous) dimension 5 couplings like $QQQL$.  Only $R_2$ in this class
allows for the presence of neutrino Majorana masses.

{\bf ii)}  The baryon triality  ${\IZ_3}$ generated by $B_3=R_3L_3$. This allows all dim 4 L-violating couplings but 
forbids the B-violating one $UDD$. It also forbids  B/L-violating dimension 5 operators like $QQQL$
but allows for Majorana neutrino masses. 

{\bf iii)} The lepton triality $L_3$ which forbids dimension  4 and 5 L-violating couplings. On the other hand it forbids
neutrino Majorana masses.

{\bf iv)} The $R_3L_3^2$ symmetry which forbids all dangerous dimension 4 and 5 B/L violating operators but
again does not allow for neutrino Majorana masses.

Note that all these symmetries forbid baryon decay through dimension four operators.
Among these symmetries only R-parity and baryon triality (or hexality, which is the product of both) allow for
neutrino Majorana masses and hence are phenomenologically preferred.
In addition only baryon triality (or hexality) forbid dimension 5 B/L-violating 
operators.  It is worth to notice that if SUSY is found with L-violating $QDL$ couplings at LHC, it would be
evidence for baryon triality and non-unification since we have shown that $SU(5)$ unification may only be
consistent with $R_N$ discrete symmetries like R-parity. 

Given their important role in models with underlying $SU(5)$GUTs, we have further studied the realization of $R_N$ discrete symmetries in F-theory models with split $S(U(4)\times U(1))$ spectral cover construction. A more systematic understanding of $U(1)$ symmetries and their discrete subgroups in other F-theory setups is an interesting new direction. 

Finally, we have further explored the realization of the $\IZ_2$ R-parity from several different sources, including the constraints from cancellation of K-theory charge, and from the existence of instanton sectors with minimal instanton number 2 (due to their $Sp$-type orientifold projection)

In our opinion the source of discrete gauge symmetries described in this paper provide us with
the best available understanding of proton stability in the MSSM. We expect further progress in extending the study of discrete gauge symmetries in brane models (including e.g. R-symmetries, non-abelian symmetries, etc), and systematically studying other related setups (like general F-theory local GUTs, M-theory models, etc).

\begin{center}
{\bf\large Acknowledgements}
\end{center}

We thank F. Marchesano, E. Palti and G. Shiu for useful discussions.
This work has been partially supported by the grants FPA 2009-09017, FPA 2009-07908, Consolider-CPAN (CSD2007-00042) from the MICINN, HEPHACOS-S2009/ESP1473 from the C.A. de Madrid and the contract ``UNILHC" PITN-GA-2009-237920 of the European Commission. M.B-G. acknowledges the finantial support of the FPU grant AP2009-0327. P.S. acknowledges support by the CSIC grant JAE-Pre-0800401 and would like to thank the HKIAS for hospitality during the early stages of this work. A.M.U. thanks M. Gonz\'alez for encouragement and support.

\newpage

\appendix

\section{Tilted orientifolds}
\label{ap:tilted-orientifolds}

Before the orientifold projection we can introduce a basis of 3-cycles, $\{\tilde\a_k\},\{\tilde \b_k\}$, satisfying $\tilde\a_k\cdot \tilde\b_{l}=\delta_{kl}$. In the main text we have focused on the situation where $\tilde\a_k\to \tilde\a_k$ and $\tilde\b_k\to-\tilde\b_k$, which were denoted $\alpha_k$ and $\beta_k$. The orientifold action is however compatible with other possibilities, e.g. in which for a subset of $k$'s we have $\tilde\a_k\to \tilde\a_k-\tilde\b_k$, $\tilde\b_k\to -\tilde\b_k$, or in which for a subset we have $\tilde\a_k\to \tilde\a_k$, $\tilde\b_k\to -\tilde\b_k+ \tilde\a_k$. This kind of situation is familiar in compactification with tilted $\IT^2$'s, so we dub them `tilted orientifolds'. Actually, the latter turns into the former possibility by renaming $\tilde\a'=2\tilde\a-\tilde\b$, $\tilde\b'=\tilde\a$, so we focus on the action $\tilde\a\to \tilde\a-\tilde\b$, $\tilde\b\to -\tilde\b$ for concreteness. For simplicity, we also assume that this happe
 ns for all $k$. 

The cycles with definite parity are given by $\alpha_k=2\tilde\a_k-\tilde\b_k$, $\beta_k=\tilde\b_k$, and $\alpha_k\cdot \beta_l=2\delta_{kl}$. The 3-cycle wrapped by the D6$_A$-branes and their images are
\beqa
[\Pi_A]\, =\, r_A^k \, \tilde\a_k\, +\,s_A^k\, \tilde\b_k\, =\, \frac 12 r_A^k \alpha_k \, +\, {\tilde s}_A^k\beta_k  \quad ,\quad [\Pi_{A'}]\, =\,  \frac 12 r_A^k \alpha_k \, -\, {\tilde s}_A^k\beta_k
\eeqa
 where we have introduced ${\tilde s}_A^k=s_A^k+\frac 12 r_A^k$. We define the RR 2-forms and scalars as
 \beqa
 B_k\, =\,  \int_{\beta_k} C_5 \quad , \quad a_k=\int_{\alpha_k} C_3
 \eeqa
Notice that since $\alpha_k=2\tilde\a_k-\tilde\b_k$, there is a factor of 2 in the duality relation; this will shortly play a role in the discussion. The $BF$ couplings read
 \beqa
N_A \sum_k {\tilde s}_A^k\, B_k\wedge F_A
 \eeqa
 Considering a $U(1)$ linear combination $Q=\sum_A c_A Q_A$, under a $U(1)$ gauge transformation, the shift in the RR scalar is
\beqa
A_\mu\to A_\mu+\partial_\mu \lambda \quad ,\quad a_k \to a_k\,+\, 2\,\sum_A c_A N_A {\tilde s}_A^k\lambda
\eeqa
where the factor of 2 (related to the above mentioned one) arises because $\alpha_k=2\tilde\a_k-\tilde\b_k$; note that this ensures the coefficient to be integer, even though the ${\tilde s}$ can be $\frac 12$ (mod $\IZ$). Noting that $[\Pi_A]\cdot[\alpha_k]=-2{\tilde s}_A^k$, the condition for a $\IZ_n$ subgroup to remain as a discrete gauge symmetry is given by the expression (\ref{zn-condition}).

\section{$\IZ_2$ symmetries and R-parity from $Sp(2)$ instantons}
\label{ap:sp}

As mentioned at the end of section \ref{sec:instanton-effects}
it is possible to construct models in which  a $\IZ_2$ subgroup of {\em each} $U(1)$ is automatically preserved by instantons. This happens in compactifications for which all D-brane instantons mapped to themselves under the orientifold action (invariant instantons, for short) have  $USp$ worldvolume symmetry (rather than $SO$). More specifically, a $U(1)$ in the model with associated homology charge $[\Pi_Q]$ is violated by a D2-brane instanton (with Chan-Paton multiplicity $k$) on $\Pi_{\rm inst}$ by an amount
\beqa
&k [\Pi_Q]\cdot \bigl(\, [\Pi_{\rm inst}]\, -\, [\Pi_{\rm inst}']\, \bigr) \in 2\IZ \quad \mbox{for non-invariant instantons}\nonumber \\
& k[\Pi_Q]\cdot [\Pi_{\rm inst}] \in 2\IZ \quad \mbox{for invariant instantons}
\eeqa
where the first line corresponds to an even quantity due to the contributions of branes and images, while the second is an even quantity due to the $USp$ character assumed for invariant instantons.

An example of compactification realizing this mechanism is the $\IT^6/(\IZ_2\times \IZ_2)$ orientifold with positively charged O6-planes. In the usual choice with negative charge for all four kinds of  O6-planes,  invariant D6-branes have $USp$-type orientifold projection, so invariant instantons have $SO$-type projections; this is reversed in the compactification with positively charged O6-planes. The models are necessarily non-supersymmetric, since RR tadpole cancellation must be achieved by the introduction of (possibly intersecting) anti-D6-branes. However, we expect that more general CY orientifold compactification may  allow the realization of this mechanism in a supersymmetric fashion.

These $\IZ_2$ symmetries are actual gauge discrete symmetries of the theory, and with some massage they can be made manifest in terms of $BF$ couplings\footnote{The extra factor of 2 in the discrete symmetries in the tilted orientifolds of appendix \ref{ap:tilted-orientifolds} can also be transferred into the coefficient of the $BF$ coupling by a similar argument.}. Following \cite{Banks:2010zn}, a lagrangian description for a $\IZ_n$ gauge theory is
\beqa
t^2(da-nA)\wedge *(da-nA)\, +\, \frac 12 F\wedge *F
\label{higgs-picture}
\eeqa
where the order of the symmetry is given by $n$, if the scalar $a$ has periodicity $2\pi$, and charges under $A$ are integer. Dualization of the above scalar to $*da=dB\equiv H$ leads to the formulation in terms of $BF$ couplings
\beqa
\frac{1}{(4\pi)^2t^2}\, H\wedge *H\, +\,\frac{in}{2\pi} B\wedge F\,+\, \frac 12F\wedge *F
\label{bf-picture}
\eeqa
in which the discrete symmetry is identified by the coefficient $n$, and the $t$-dependent factor is not relevant. In the above context, in which all invariant instantons are of $USp$-type, instanton numbers are effectively truncated to be even. The periodicity of the RR scalar $a$ is halved to $\pi$, so we must introduce a scalar $a'=2a$, and write the first term in (\ref{higgs-picture}) as
\beqa
t^2(da-nA)\wedge *(da-nA)\, =\, \left(\frac t2\right)^2(da'-2nA)\wedge *(da'-2nA)\, 
\eeqa
The latter expression shows that the actual gauge symmetry is $\IZ_{2n}$. Equivalently, we recover the structure (\ref{bf-picture}) with $BF$ coupling coefficient $2n$ (and an irrelevant change $t\to t/2$). So, even for $n=1$ there is a $\IZ_2$ discrete gauge symmetry associated to the restriction in the available instanton numbers; models with only $USp$-type instanton provide a microscopic implementation of the phenomenon in \cite{Seiberg:2010qd} (see also \cite{Hellerman:2010fv}).

Obtaining a $\IZ_2$ subgroup of every single $U(1)$ in the model may however not be necessarily appealing from the phenomenological viewpoint. For instance, they may well prevent some instantons from generating phenomenologically interesting couplings. A more economic and better targeted possibility is to consider models where a $\IZ_2$ subgroup of {\em some} $U(1)$ is preserved, because all invariant instantons violating it have $USp$ projection (whereas others, not violating the $U(1)$, may have $O(1)$ projections). 

As an example, we consider a version of the model in table \ref{newfamily}, embedded in a $\IT^6/(\IZ_2\times \IZ_2)$ orientifold; as usual, this requires doubling the number of D6-branes in each stack to $2N$ to generate an $U(N)$ symmetry. We make the usual choice of discrete torsion corresponding to Hodge numbers $(h_{11},h_{21})=(51,3)$. As shown in \cite{Angelantonj:1999ms}, this choice requires having an {\em even} number of negatively charged O6-planes, among the four kinds present in the model\footnote{The opposite choice of discrete torsion requires an odd number of negatively charged O6-planes.}; rather than the usual choice of choosing all O6-planes to have negative charge, we choose negatively charged O6-planes along $[a_1][a_2][a_3]$, $[a_1][b_2][b_3]$ and positively charged O6-planes along $[b_1][a_2][a_3]$ and $[b_1][b_2][b_3]$.  This does not modify the appearance of the SM spectrum from the visible branes, since the orientifold signs only enter in the multipl
 icities of
 two-index tensor representations in $AA'$ sectors, which are massless in the model, and would only change the set of hidden branes to cancel the tadpole, ignored for simplicity. Consider the $U(1)_c$ gauge factor, which as shown in Section \ref{sp-class} is associated to the $R_N$  generators. The only invariant instanton intersecting $[\Pi_c]$ is that wrapped on $[b_1][b_2][a_3]$, and it has $Sp$ orientifold projection due to our choice of O6-plane charge. Hence a $\IZ_2$ subgroup of $U(1)_c$, corresponding to R-parity, is automatically preserved by all D-brane instanton effects in the model.

The interpretation of these symmetries as gauge discrete symmetries in terms of $BF$ couplings works in complete analogy with the above case.

\end{document}